 [2005/06/22 5.2/AAS markup document class]%
\documentclass[preprint]{aastex}%
\usepackage{multirow}
 \usepackage{url}

\makeindex

\makeatletter 

\let\o@verbatim\verbatim
\def\verbatim{%
  \ifhmode\unskip\par\fi
  \ifx\@currsize\normalsize
     \small
  \fi
  \o@verbatim
}

\renewcommand \verbatim@font {%
  \normalfont \ttfamily
  \catcode`\<=\active
  \catcode`\>=\active
}

\RequirePackage{shortvrb}
\MakeShortVerb{\|}

\begingroup
  \catcode`\<=\active
  \catcode`\>=\active
  \gdef<{\@ifnextchar<\@lt\@meta}
  \gdef>{\@ifnextchar>\@gt\@gtr@err}
  \gdef\@meta#1>{\m{#1}}
  \gdef\@lt<{\char`\<}
  \gdef\@gt>{\char`\>}
\endgroup
\def\@gtr@err{%
   \ClassError{ltxguide}{%
      Isolated \protect>%
   }{%
      In this document class, \protect<...\protect>
      is used to indicate a parameter.\MessageBreak
      I've just found a \protect> on its own.
      Perhaps you meant to type \protect>\protect>?
   }%
}
\def\verbatim@nolig@list{\do\`\do\,\do\'\do\-}

\newcommand{\m}[1]{\mbox{$\langle$\it #1\/$\rangle$}}

\def\cmd#1{\cs{\expandafter\cmd@to@cs\string#1}}
\def\cmd@to@cs#1#2{\char\number`#2\relax}
\DeclareRobustCommand\cs[1]{\texttt{\char`\\#1}}
\def\GetFileInfo#1{%
  \def\filename{#1}%
  \def\@tempb##1 ##2 ##3\relax##4\relax{%
    \def\filedate{##1}%
    \def\fileversion{##2}%
    \def\fileinfo{##3}}%
  \edef\@tempa{\csname ver@#1\endcsname}%
  \expandafter\@tempb\@tempa\relax? ? \relax\relax}

\makeatother

\title{%
  Joint {\sl XMM-Newton} and {\sl Chandra} Observations of the NGC~1407/1400 Complex: \\
  a Tail of an Early-Type Galaxy and a Tale of a Nearby Merging Group 
}%

\author{Yuanyuan Su\altaffilmark{1}, Liyi Gu\altaffilmark{2}, Raymond E. White III\altaffilmark{3} and Jimmy Irwin\altaffilmark{3}}

\altaffiltext{1}{Department of Physics and Astronomy, University of 
California at Irvine, 4129 Frederick Reines Hall, Irvine, CA 92697, USA}

\altaffiltext{2}{Department of Physics, University of 
Tokyo, 7-3-1 Hongo, Bunkyo-ku, Tokyo, Japan}

\altaffiltext{3}{Department of Physics and Astronomy, University of 
Alabama, Box 870324, Tuscaloosa, AL 35487, USA}

\email{yuanyuas@uci.edu}

\begin{document}

\expandafter\GetFileInfo\expandafter{\jobname.tex}%

\begin{abstract}
The nearby group centered on its bright central galaxy NGC~1407 has been suggested to be an unusually dark system from previous kinematic studies. It is also known for hosting a bright galaxy, NGC~1400, with a large radial velocity (1200 km~s$^{-1}$) with respect to the group center. 
Previous {\sl ROSAT} X-ray observations revealed an extended region of enhanced surface brightness just eastward of NGC~1400.
We investigate the NGC~1407/1400 complex with {\sl XMM-Newton} and {\sl Chandra} observations. 
We find that the temperature and metallicity of the enhanced region
are different (cooler and more metal rich) than those
of the surrounding group gas, but consistent with
those of the ISM in NGC~1400.
The relative velocity of NGC~1400 is large enough that
much of its ISM could have been ram pressure stripped
while plunging through the group atmosphere.
We conclude that the enhanced region is likely to be hot gas stripped from the ISM of NGC~1400.
We constrain the motion of NGC~1400 using the 
the pressure jump at its associated stagnation front and the total mass profile of the NGC~1407 group.
We conclude that NGC~1400 is moving within $\sim30^{\circ}$ of the line-of-sight with Mach number $\mathcal{M}\lesssim3$.
We do not detect any obvious shock features in this complex, 
perhaps due to the highly line-of-sight motion of NGC~1400.
With an {\sl XMM-Newton} pointing on the relatively relaxed eastern side of NGC~1407, we derive a hydrostatic mass for this group of $\sim1\times 10^{13}$ $M_\odot$ within 100 kpc. The  total mass extrapolated to the virial radius (681 kpc) is 3.8$\times 10^{13}$ $M_\odot$, which puts an upper limit of $\sim$300 $M_\odot/L_{B_\odot}$ on the mass-to-light ratio of this group. This suggests that the NGC~1407 group is {\sl not} an unusually dark group. 
\end{abstract}


\section{\bf Introduction}
\smallskip

The nearby group centered on the elliptical galaxy NGC~1407 is
notable because it appears to contain a bright lenticular galaxy, NGC 1400,
with a very large velocity (1200 km s$^{-1}$)  relative to NGC~1407.
This has led to suggestions that the NGC 1407 group is one of the darkest known galaxy systems
(Gould 1993).
This group belongs to the Eridanus supergroup, which contains three main groups: 
the NGC~1407 group, the NGC~1332 group and the Eridanus group (Brough et al.\ 2006). 
Among them the NGC~1407 group is the most dynamically and morphologically relaxed and
NGC~1407 is the brightest galaxy in the supergroup.
The NGC~1407 group has relatively relaxed X-ray morphology and has a cool core centered on NGC~1407 
(Zhang et al.\ 2007).
NGC~1407 is X-ray luminous ($L_X=8.6\times10^{40}$ erg s$^{-1}$ in 0.1--2.0 keV, within two optical 
effective radii; Su \& Irwin 2013), with evidence for recurrent radio outbursts (Giacintucci et al.\ 2012).
The Eridanus supergroup may collapse into a massive cluster in the future, with  NGC~1407  as its 
future brightest cluster galaxy (BCG). 

Hierarchical structure formation theory predicts an upper limit for the total $M/L$ of galaxy systems, 
assuming high mass systems are formed from low mass systems with similar
stellar mass fractions (Kauffmann et al.\ 1999; Marinoni \& Hudson 2002). 
The observational discovery of galaxy groups and clusters with extremely small baryonic mass fractions 
may indicate a gap in our knowledge about the current structure formation paradigm
(Balogh et al.\ 2008; Bower at al.\ 2006; Giodini et al.\ 2009; Lin et al.\ 2003).
The NGC~1407 group is a possible example of such a system.

Romanowsky et al.\ (2009) found a very low baryon mass fraction $f_b$ in the NGC 1407 group,
after using globular cluster kinematics (within 60 kpc of NGC~1407) to measure the group mass; 
extrapolating their results to the virial radius, they found
$M/L \sim800$ $M_\odot/L_{B_\odot}$ and $f_b \sim 0.004$, the latter being 
much lower than the cosmological value of $f_b=0.17$ (Hinshaw et al.\ 2009). 
Zhang et al.\ (2007) studied the hot gas of this group with {\sl Chandra} ACIS-S
X-ray observations centered on NGC~1407, as well as {\sl ROSAT} observations
extending out to 75 kpc from NGC~1407.
Assuming hydrostatic equilibrium, they inferred a smaller value of $M/L=311\pm60$ $M_\odot/L_{B_\odot}$ within its virial radius.  
The large velocity difference places NGC~1400's group membership in doubt and
casts more doubt on the total mass (hence, mass-to-light ratio) estimates for this group.  
Brough et al.\ (2006) studied the galaxy dynamics of this group and derived the total mass by both excluding and including NGC~1400 and five other galaxies, which yielded low and high $M/L$ estimates of 600 and 1200 $M_\odot/L_{B\odot}$, respectively.

In estimating masses of galaxy groups, both  X-ray techniques and dynamical probes have their drawbacks. 
X-ray techniques assuming hydrodynamic equilibrium may be affected by group gas being
out of equilibrium, having significant 
non-thermal pressure support, and/or containing multi-phase gas.
Dynamical analyses tend to lack the spatial extent of X-ray observations and require
assumptions about velocity distributions.
Despite the fact that there is no general consensus about the $M/L$ of the NGC~1407 group, 
all estimates to date are above the typical range for groups, 
which is $M/L_B \sim 60-300$ $M_\odot/L_{B_\odot}$ (Eke et al.\ 2006).
In this paper, we reveal
 some thermal substructures in the gas associated with NGC~1400, which undermines the assumption of hydrostatic equilibrium for X-ray analyses.  
This motivated us to propose an {\sl XMM-Newton} pointing on the eastern side of the NGC~1407 group out to 100 kpc, where the thermal structure is relatively relaxed, in order to better estimate the total mass of this possibly dark system through hydrostatic X-ray techniques.

\begin{figure} 
\epsscale{1.05}
\plotone{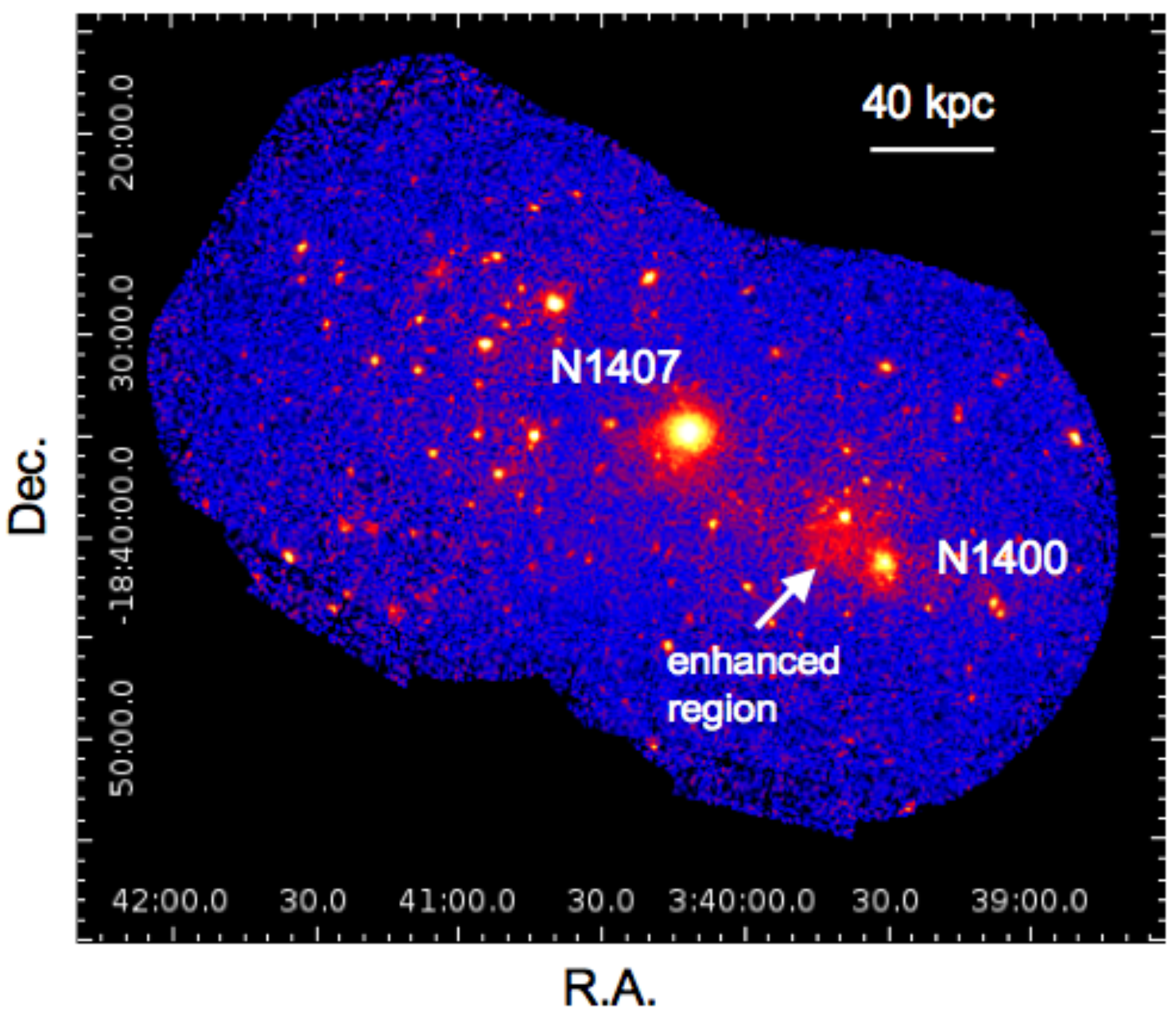}
\caption{\label{fig:1407} {\sl XMM-Newton} image (0.5-2.0 keV) of the NGC~1407/1400 complex. [{\sl see the electronic edition of the journal for a color version of this figure.}]}
\end{figure}

We also analyzed the existing archival {\sl XMM-Newton} pointing on the western side of NGC~1407, which covers both NGC~1407 and NGC~1400.
{\sl XMM-Newton} imaging reveals an extended region of enhanced surface brightness just east of NGC~1400 (Figure~\ref{fig:1407}), which was attributed to be a background object in the previous {\sl ROSAT} study (Zhang et al.\ 2007).
We suspect that this enhanced region is 
hot gas stripped out of NGC~1400 due to its large relative velocity through the group. 
Furthermore, we found an apparently  heated region between NGC~1407 and NGC~1400, 
potentially an indicator of an ongoing collision. 
An additional {\it XMM-Newton} pointing covering the eastern side of NGC~1407 provides us an azimuthally complete coverage. 
This helps us to diagnose whether the apparently heated region between NGC~1400 and NGC~1407 is indeed 
non-gravitational heating or simply a reflection of the gravitational well of the NGC~1407 group. 
To further unveil the dynamical state of this complex,
we also acquired an additional {\sl Chandra} ACIS-S pointing covering NGC~1400 and the enhanced region, 
which allows us to infer the gas dynamics of the enhanced region.
In this paper, we present these observations and discuss their implications.

We assume $H_0 = 70$ km~s$^{-1}$ Mpc$^{-1}$, $\Omega_{\Lambda}=0.7$, and  $\Omega_M=0.3$.  
The central  dominant galaxy NGC 1407
resides at a luminosity distance of  $D_L$ = 22.08 Mpc ($1^{\prime}$ = 6.5 kpc; 
NED\footnote{\url{http://ned.ipac.caltech.edu}}). 
Throughout this paper, all uncertainties are given at the $90\%$ confidence level unless otherwise stated.  
Observations and data reduction are described in \S2. We report our results in \S3 and
describe our proposed scenario in \S4.
We summarize our main conclusions in \S5. 
\bigskip

\medskip
    
\section{\bf Observations and data reduction} 
\smallskip


The NGC~1407 group was previously observed on its western side with {\sl XMM-Newton}, with NGC~1400 in the field of view. We proposed an {\sl XMM-Newton} observation on the eastern side of this group, away from NGC~1400. A mosaic image of these two pointings is shown in Figure~\ref{fig:1407}. Observation logs are listed in Table~1.   
Only data from the MOS and PN detectors of the European Photon Imaging Camera (EPIC) are reported in this paper. 
The standard {\sl Science Analysis System} ({\sl SAS} 11.0.0)\footnote{\url{http://xmm.esac.esa.int/sas/}} pipeline 
tools were used throughout the analyses. The tasks {\tt emchain} and {\tt epchain} were used to generate calibrated 
event files from the raw data. 
MOS events were selected with event parameter 
{\tt PATTERN} $\leq12$, while {\tt PATTERN} $\leq 4$ was used to select PN events. 
The removal of bright pixels and hot columns was done by applying the expression 
(FLAG==0). 
Point sources resolved with {\tt edetect\_chain} and verified by eye were removed. To remove the soft proton flares from the source data, we used the {\tt mos-filter} and {\tt pn-filter} tasks in {\sl XMM-ESAS}. We also extracted images in the soft energy band (0.3--1.0 keV) of each pointing and detector to make sure  there were no remaining soft proton flares on any CCD chip.
The remaining exposure times, after filtering for background flares, are shown in Table~1. 

We also observed NGC~1400 and the region of enhanced surface brightness with {\sl Chandra} ACIS-S.
The observation log is shown in Table~1. 
{\sl CIAO}~4.5 and {\sl CALDB}~4.5.5\footnote{\url{http://cxc.harvard.edu/ciao/index.html}} were used throughout the analysis.
Data were reprocessed from level 1 events, which guarantees the use of updated and consistent calibrations.  
Only events with grades 0, 2, 3, 4, and 6 were included. We also removed bad pixels, bad columns, and node boundaries.
We filtered background flares with the light curve filtering script {\tt lc\_clean}.  The effective exposure times are shown in Table~1.

\subsection{\sl Image analysis}

\subsubsection{\sl XMM-Newton}
We extracted {\sl XMM-Newton} images in the 0.5--2.0 keV energy band for each pointing and detector from the processed and filtered events files.  We generated their exposure maps with the {\sl SAS} task {\tt exposure}, which also corrects for vignetting.
We divided each image by its associated exposure map. For PN images, we also performed out-of-time corrections. 
Exposure- and vignetting-corrected images of MOS1, MOS2 and PN were added together for both the western and eastern pointings. Images were smoothed to achieve a signal-to-noise ratio of 40. 
A final mosaic image (0.5--2 keV) for both western and eastern pointings is shown in Figure~\ref{fig:1407}.

\subsubsection{\sl Chandra} 

We extracted {\sl Chandra} images in three energy bands: 0.5--2.0 keV, 0.5--1.5 keV and 1.5--3.0 keV. Bright point sources resolved with {\tt wavdetect} were removed and we divided these images by their exposure maps. 
We adopted ``blank-sky" observations\footnote{\url{http://cxc.harvard.edu/contrib/maxim/acisbg/}} to estimate the background components. 
We scaled the background level to the source data using the hard band (10--12 keV) count rate ratios between the source data and blank sky, 
and subtracted the background component from each image.
The final images were smoothed with a Gaussian kernel. The {\sl Chandra} image in the 0.5--2.0 keV band is shown in Figure~\ref{fig:sur}. 

\subsection{\sl Spectral analysis with XMM-Newton}

\begin{figure} 
\epsscale{0.8}
\plotone{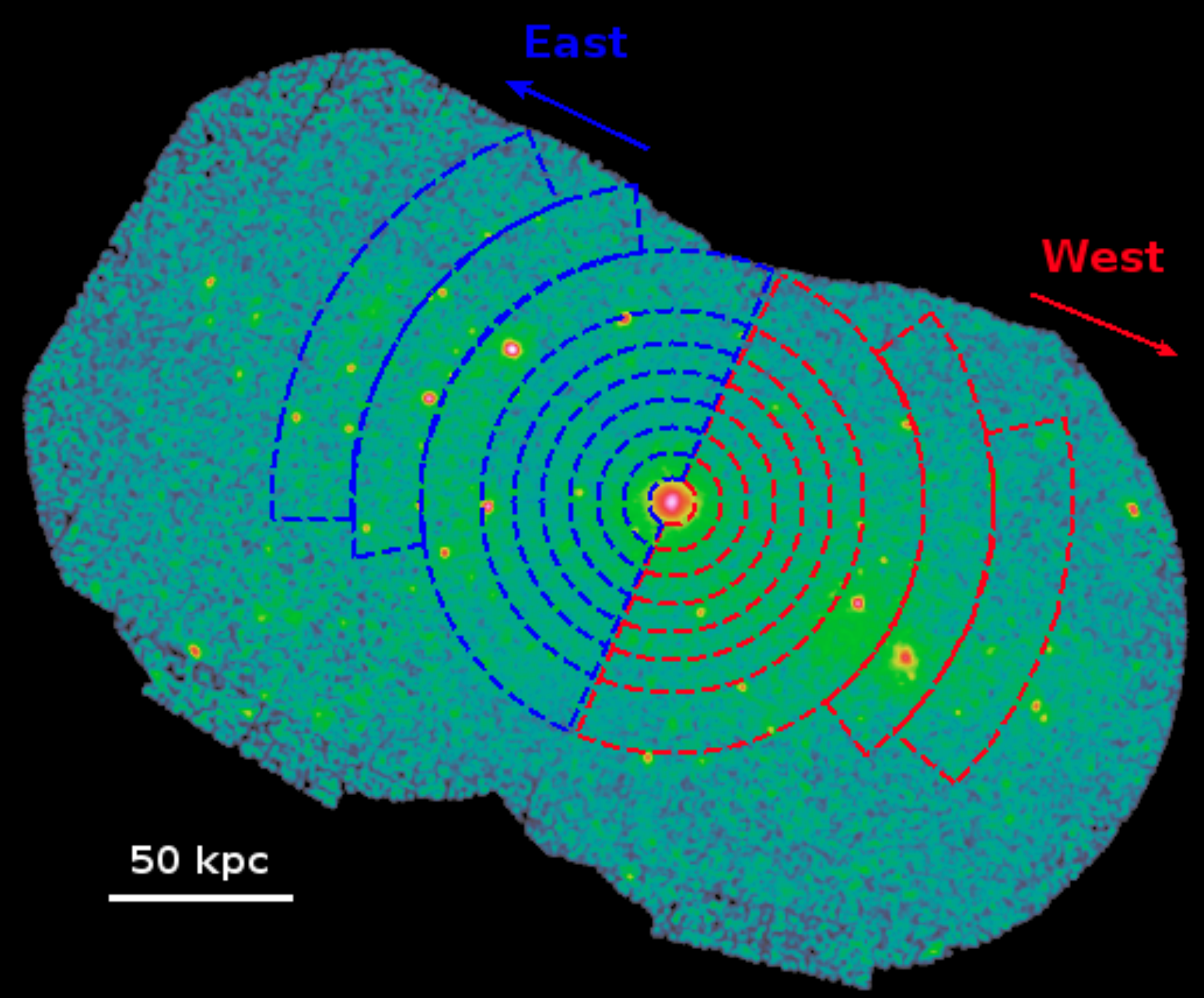}
\caption{\label{fig:14072}  {\sl XMM-Newton} image of the NGC~1407/1400 complex (0.5-2.0 keV band). 
Regions  extracted for spectral analysis in eastern and western directions are indicated. [{\sl see the electronic edition of the journal for a color version of this figure.}]}
\end{figure}

We extracted {\sl XMM-Newton} spectra from 10 hemispherical annular sections for each of the western and eastern pointings, centered on NGC~1407 and extending westward and eastward.
Annular regions 1--10 on each side have  radial ranges of
$0^\prime$-$1^\prime$, $1^\prime$-$2^\prime$, $2^\prime$-$3.15^\prime$, 
$3.15^\prime$-$4.3^\prime$, $4.3^\prime$-$5.45^\prime$, $5.45^\prime$-$6.6^\prime$, $6.6^\prime$-$8^\prime$, $8^\prime$-$10.5^\prime$, $10.5^\prime$-$13.3^\prime$ and $13.3^\prime$-$16.7^\prime$, respectively,
as indicated in Figure~\ref{fig:14072}.  
 Their associated background spectra are extracted from Filter Wheel Closed (FWC) data\footnote{\url{http://xmm2.esac.esa.int/external/xmm_sw_cal/background/filter_closed/}} for each region and detector (see below). We adopted C-statistics for all our spectral analyses and  grouped spectra to have at least one count per bin.
 Redistribution matrix files 
(RMFs) were generated for each region and detector using the {\sl SAS} task {\tt rmfgen}; 
 ancillary response files 
(ARFs) for each region and detector were produced with the task {\tt arfgen}.
We adopted the solar abundance standard of Asplund (2009) in thermal spectral models. 
Energy bands were restricted to 0.3--7.0 keV for the MOS CCDs and 0.5--7.0 keV for the PN CCDs, 
where the responses are best calibrated.

\subsubsection{\sl Background determination} 

We determined the particle background with FWC data. 
To minimize the effect of instrumental variations over time, 
the FWC data were selected from observations at similar times as the western and eastern {\sl XMM-Newton} pointings. 
The exposure times of FWC data were chosen to be at least 30 ksec to ensure a sufficient signal-to-noise ratio.  
We processed and filtered the FWC event files in the same way as we did the {\sl XMM-Newton} observations of this group. 
The FWC event files were then cast onto the same WCS positions and angles as the associated source data. 
We obtained the 0.3--12 keV count rates in the unexposed chip corners for both the FWC data and source data for each pointing. 
We scaled the backgrounds
of all the FWC spectra by these count rate ratios (via the parameter keyword {\tt BACKSCAL}) 
before spectral analysis.

We used {\sl XSPEC} to fit the spectra from regions 3--10 (2$^{\prime}$--16.7$^{\prime}$) of both pointings simultaneously.
 We did not include regions central 1 and 2 because the contribution from low mass X-ray binaries (LMXBs) in NGC~1407 may be degenerate with the cosmic X-ray background (CXB) component. In addition, the central regions are more dominated by group emission.  
The spectral model we adopted consists of the group emission and the X-ray background. 
We used an {\tt apec} thermal emission model to represent the group emission ({\tt apec$_{\rm group}$}),
with a redshift fixed at 0.0057 (NED). 
Our X-ray background model consists
of an {\tt apec} thermal emission model for the Local Bubble 
({\tt apec$_{\rm LB}$}), 
an {\tt apec} thermal emission model for the Milky Way ({\tt apec$_{\rm MW}$})  
in the line of sight (Smith et al.\ 2001), 
and a power law with an index of 1.41 ({\tt pow$_{\rm CXB}$}) representing the CXB
(De Luca \& Molendi 2004). 
All these components but the Local Bubble were assumed to be absorbed
by foreground (Galactic) cool gas, with the absorption characterized by 
the {\tt phabs} photoelectric absorption model.  
Photoionization cross-sections were from Balucinska-Church \& McCammon (1992). 
We adopted a Galactic 
hydrogen column of $N_H=5.4\times10^{20}$ cm$^{-2}$ toward NGC~1407, deduced 
from the Dickey and Lockman (1990) map incorporated in the {\sl HEASARC}
$N_H$ tool.
The temperature, abundance and normalization of the {\tt apec} model for the group emission were allowed to vary independently within each region.  
The temperatures of the Milky Way and Local Bubble
components were fixed at 0.25 keV and 0.10 keV, 
respectively (Kuntz \& Snowden 2000);
the abundance and redshift of these two local components were fixed at solar and 0, respectively, in their 
corresponding {\tt apec} thermal emission models. 
The normalizations of the X-ray background were linked to each other by their areas to ensure that
each component had the same surface brightness in all  regions. 
The final spectral fitting model was
\begin{equation}
{\tt apec}_{\rm LB}+{\tt phabs} \times ({\tt apec}_{\rm MW}+{\tt pow_{\rm CXB}}+{\tt apec}_{\rm group}).
\end{equation}


The best-fit surface brightness for each X-ray background component is shown in Table~2. 
We performed Markov chain Monte Carlo (MCMC) simulations
for the spectral fits. With the {\tt margin} command in XSPEC, we marginalized over the uncertainties of the normalization of each X-ray background component (CXB, MW and LB). Marginalized values are also shown in Table~2.  

\subsubsection{\sl Projected spectral analysis of group emission}

We fit the individual projected spectra of regions 1--10, for both western and eastern pointings, 
with the model shown in (1).
For regions 1--2 only, we added an additional power law component  with an index of 1.6 to represent unresolved LMXBs (Irwin et al.\ 2003)
in the central galaxy NGC~1407.   
All background components were fixed at the surface brightness values 
determined earlier.  
We obtained the best-fit projected temperature profiles for both western and eastern pointings as shown in Figure~\ref{fig:tp} and listed in Table~3; temperatures range from $0.8-1.3$ keV.

\subsubsection{\sl Deprojected spectral analysis of group emission}

\begin{figure} 
\epsscale{0.9}
\plotone{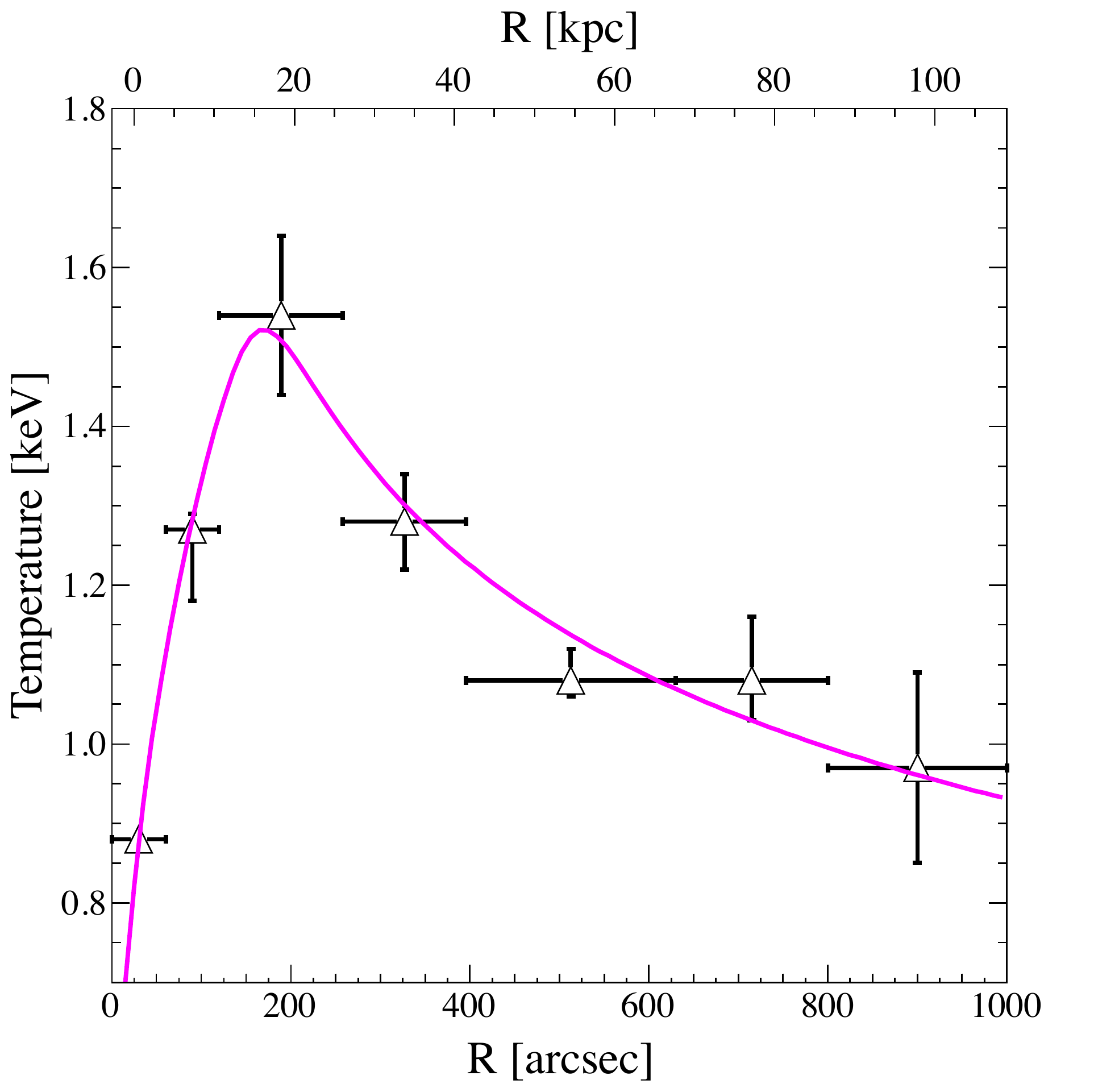}
\caption{\label{fig:t} Deprojected temperature profile centered on NGC~1407, extending eastward. 
Magenta line: best-fit temperature profile of Vikhlinin et al.\ (2006). [{\sl see the electronic edition of the journal for a color version of this figure.}]}
\end{figure}

\begin{figure} 
\epsscale{0.9}
\plotone{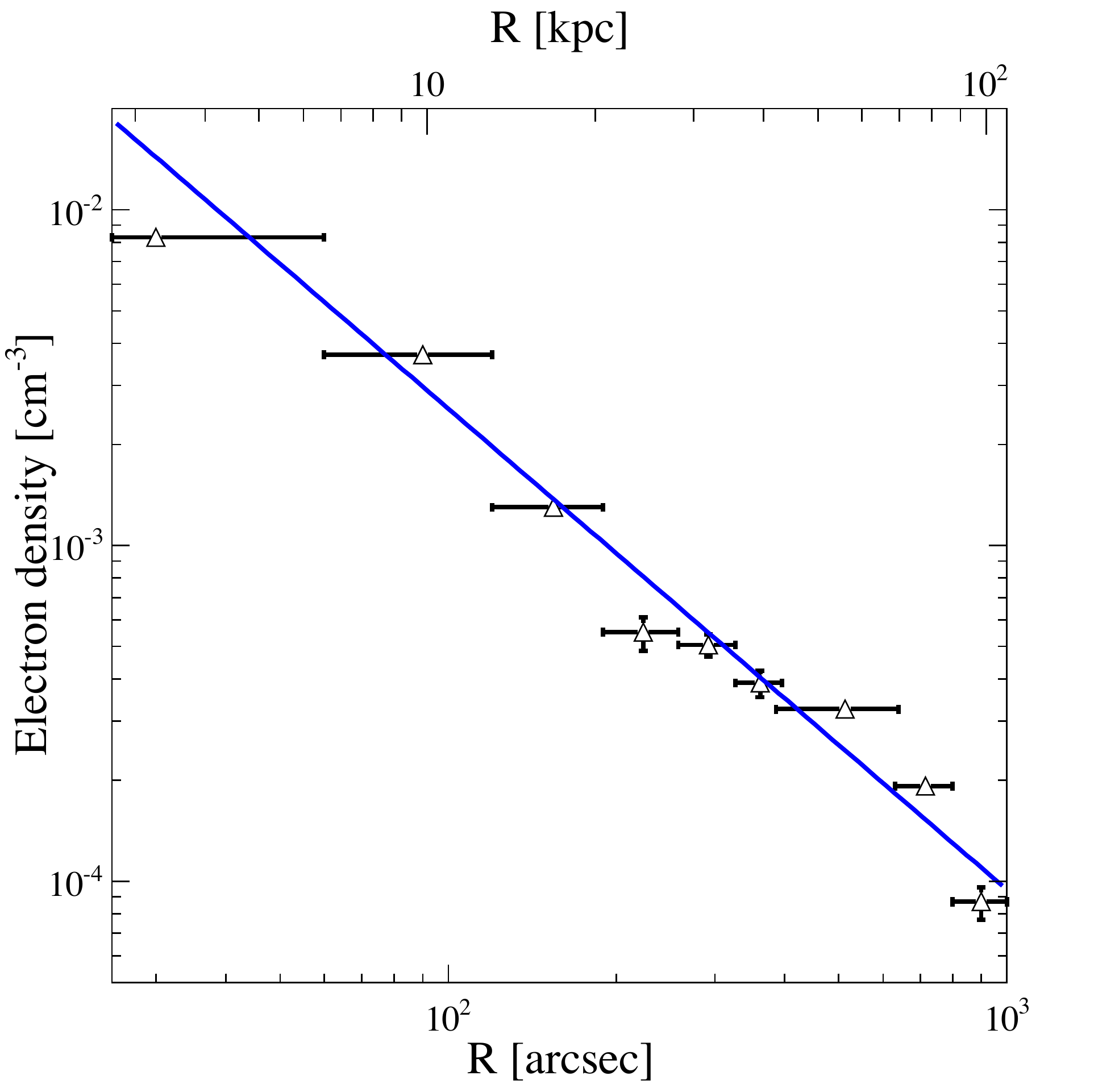}
\caption{\label{fig:ne} Deprojected electron density profile centered on NGC~1407, extending eastward. 
Blue line: best-fit single $\beta$-model profile. 
[{\sl see the electronic edition of the journal for a color version of this figure.}]}
\end{figure}

We also performed a deprojected spectral analysis 
for regions 1--10 of the eastern pointing only, since its thermal structure is less disturbed, without obvious substructure. 
We want to deproject only the group emission and
{\sl XSPEC v12} allows us to fit the group and X-ray background emission separately with different models.
We characterized the group emission with a model of the form: 
$${\tt projct}\times[{\tt phabs}\times({\tt apec_{\rm group}}+\tt pow_{1.6})],$$
where the {\tt projct} model projects 3-D shells into 2-D annuli with the same center;
the {\tt pow$_{1.6}$} model represents the LMXB component, with variable
normalizations in (the central) regions 1--2 and zero normalization 
in regions 3--10 outside the center.   
The X-ray background emission was characterized by a second (unprojected) model:
$${\tt apec}_{\rm LB}+{\tt phabs}\times({\tt apec}_{\rm MW}+{\tt pow_{\rm CXB}}),$$
with components described above.
These two models were fit simultaneously to the spectra. 
The surface brightness of each background component was fixed as determined earlier.  
To ensure the stability of the deprojected spectral fits, we had to link the temperatures of 
regions 3 and 4, regions 5 and 6, and regions 7 and 8. 
We also needed to 
tie the electron densities of regions 7 and 8 together.
Fitting results are listed in Table~4.
The best-fit deprojected temperature profile is shown in Figure~\ref{fig:t},
with temperatures ranging from $0.9-1.5$ keV.
The deprojected electron density of each spherical annulus
was calculated from the best-fit normalization
of the ${\tt apec}_{\rm group}$ component in each deprojected shell.
The normalization of the {\sl XSPEC} {\tt apec} model is defined as
\begin{equation} 
{\rm norm} = \frac{10^{-14}}{4\pi[D_A(1+z)]^2}\int{n_en_H}{dV}, 
\end{equation}
where 
$D_A$ is the angular distance to the group, and $V$ is the volume of the spherical annulus. 
We assume that the hot gas density has a single value in the given volume for each spherical annulus.
The deprojected electron density profile is shown in Figure~\ref{fig:ne}.

\subsection{\sl Temperature Map} 


We generated a temperature map from 
the western {\sl XMM-Newton} pointing (see Figure~\ref{fig:t1}), 
using adaptive binning methods described in detail in O'Sullivan et al.\ (2005) 
and Gu et al.\ (2009).
Within a region 140 kpc square ($20^{\prime}\times20^{\prime}$) 
in the western pointing, we defined 6000 knots randomly distributed with a
separation of $\Delta_{\rm ij}<10^{\prime\prime}$ between any two  knots $i$ and $j$.
To each knot we assigned an adaptively-sized circular cell, centered on the knot and containing 
at least 3000 counts (after background subtraction)
in the respective energy band used for each detector type. 
Typical cell radii ranged from $\sim1^{\prime}$ near the group center to $\sim2^{\prime}$ at the edge of the field of view.  
For each circular cell and detector, we generated the data spectrum, FWC background spectrum, RMF, and ARF files. 
The spectrum of each cell was fit to model (1), where the X-ray background components (${\tt apec}_{\rm LB}$, {\tt pow$_{\rm CXB}$}, ${\tt apec}_{\rm MW}$) were fixed as determined earlier. 
The uncertainties of the best-fit hot gas temperatures 
were typically $<$10\% within the central
50 kpc, increasing progressively to $\le 15\%$ at the field edge. 

For any given sky position ${\bf r}(x,y)$, we define a scale $s({\bf r})$, so that there
are at least 3000 counts contained in a circular region centered at ${\bf r}(x,y)$ within radius $s({\bf r})$. 
We calculate the temperature $T({\bf r})$ at each position ${\bf r}(x,y)$
by integrating over all the best-fit temperatures associated with each knot ${\bf r}_i$ located within the circular region:
\begin{equation}
T ({\bf r}) ={\sum\limits_{{\bf r}_i}G_{{\bf r}_i}R_{{\bf r},{\bf r}_i}T_c({{\bf r}_i}) \over {\sum\limits_{{\bf r}_i}G_{{\bf r}_i}(R_{{\bf r},{\bf r}_i})}},
~ ~ ~ ~ {\rm when}~ R_{{\bf r},{\bf r}_i} < s({\bf r}),
\end{equation}
where $R_{{\bf r},{\bf r}_i}$ is the distance from ${\bf r}(x,y)$ to ${\bf r}_i$, and $G_{{\bf r}_i}$ is the
Gaussian kernel with scale parameter $\sigma$ fixed at $s({\bf r}_{i})$.
Since $s({\bf r})$ is essentially proportional to the square root of the
local counts, the obtained temperature $T({\bf r})$ is 
less affected by the statistical uncertainties of surface
brightness fluctuations. Spatial resolutions of
$\sim10^{\prime\prime}$ are provided by the use of a compact Gaussian kernel.

\subsection {\sl Stellar mass estimates from infrared imaging}

We used $K$-band infrared imagery from the {\sl WISE} ({\sl Wide-Field Infrared Survey Explorer})
public archive to determine the stellar light distribution in the NGC~1407 group.
We analyzed a $42^{\prime} \times 42^{\prime}$ image centered on NGC~1407.
Point sources were excluded and replaced with 
locally interpolated surface brightness using {\sl CIAO} tool {\tt dmfilth}.   
In our $K$-band photometry we adopted the same radial annular regions that we used in the 
{\sl XMM-Newton} X-ray analysis (regions 1--10). 
We used an offset region on the image to determine the local infrared background surface brightness.
We subtracted this 
background surface brightness from the infrared
surface brightness in each of the areas subtended by regions 1--10.
We converted source counts in each region to a corresponding magnitude 
and corrected for Galactic extinction.
We adopted a Solar $K$-band absolute magnitude of $M_{K\odot}=3.28$ 
(Binney \& Merrifield 1998) and derived the stellar mass by adopting 
a stellar mass-to-light ratio of $M_*/L_{K}$ $=$1.17 $M_\odot/L_{K_\odot}$ 
(Longhetti \& Saracco 2009).

\section{\bf Results}
\smallskip
\medskip

\subsection{\sl NGC 1400 and the region of enhanced surface brightness}

As shown in Figure~\ref{fig:1407}, the western {\sl XMM-Newton} pointing
revealed a region with enhanced surface brightness $\sim3^\prime$ northeast of NGC~1400. 
We propose that this enhanced surface brightness may be due to hot gas stripped from the ISM of NGC~1400.
As described later, we also considered and rejected the possibility that this emission is due
to an extended background object. 
For the western {\sl XMM-Newton} pointing, we extracted spectra from a circular region centered on NGC~1400, with a radius of
 two optical effective radii ($r_{\rm e}=0.49^{\prime}=3.2~{\rm kpc}$; RC3 -- de Vaucouleurs et al.\ 1991), 
 as well as from
a circular region (radius 1.4$^{\prime}$) covering most of the enhanced region. 
We adopted local background for the analysis of these two spectral sets so  the emission from 
group gas could be subtracted out; 
we extracted the local background spectrum from a region in the eastern pointing at a
radial distance from NGC 1407 corresponding  to the average distance of 
the enhanced region and NGC~1400 from NGC 1407.
We fit the spectrum of NGC~1400 with the model 
${\tt phabs}\times({\tt apec_{\tt ISM}}+{\tt pow_{1.6}}+{\tt mekal}+{\tt pow})$, following Su \& Irwin (2013).
The thermal ISM emission is represented by 
$\tt apec_{\tt ISM}$; the ${\tt pow_{1.6}}$ powerlaw component, with an index fixed at 1.6, represents the contribution from unresolved LMXBs (Irwin et al.\ 2003);
the {\tt mekal}+{\tt pow} components represent faint stellar X-ray sources, such as cataclysmic variables (CVs) and coronally active binaries (ABs); we fixed these CV/ABs components at their estimated flux based on NGC~1400's $K$-band stellar luminosity (see Revnivtsev et al.\ 2008 for details). 

We found an ISM temperature of $0.68^{+0.06}_{-0.05}$\,keV for NGC 1400 (see Table 5).
We could not constrain the metal abundance of its ISM, so we subsequently fixed it at 0.7 $Z_\odot$, 
which is the average ISM metallicity in the early-type galaxy sample of Su \& Irwin (2013). 
We fit the spectrum of the enhanced region with a simpler model ${\tt phabs}\times{\tt apec}$, 
since there are no stellar sources in this region.  
The best-fit temperature of the enhanced region is $0.81^{+0.03}_{-0.03}$\,keV (see Table 5),
similar to the ISM temperature of NGC~1400
and somewhat cooler than adjacent group gas ($kT\approx$1\,keV).
The best-fit abundance of the enhanced region is 0.49$^{+0.33}_{-0.14}$ $Z_\odot$, which is 
twice as high as that of group gas in the eastern pointing at the same radius from NGC~1407 
as this enhanced region (0.26$^{+0.04}_{-0.07}$ $Z_\odot$; see Table 3, eastern bin from 8--10.5$^\prime$).
This abundance is consistent
with typical ISM metal abundances ($\sim0.7$ Z$_\odot$) found in early-type galaxies of this optical luminosity (Su \& Irwin 2013). 
While the temperature of the enhanced region is similar to that of the ISM of NGC~1400,
its temperature and abundance differ from those of the surrounding group gas.
This  strongly suggests 
that the enhanced surface brightness region is ISM stripped from NGC~1400.

Note that
when fitting multiple thermal components with a single temperature model, 
the best-fit metal abundance tends to be biased low (due to the so-called ``Fe-bias" Buote 2002).  
The putative stripped gas is embedded in group gas, 
so the extracted spectrum likely contains multiple temperature components. 
Unfortunately, the current data  do not allow us to constrain such 
two-temperature gas models, as was done in Su \& Irwin (2013) for a sample of early-type galaxies.
Thus, the current $\sim$ 0.5 Z$_{\odot}$ abundance result should be taken as a lower limit to
the hot gas metallicity of the enhanced region. 
This is still consistent with the enhanced surface brightness region being ISM
stripped from NGC~1400. 

We determined 0.1--2.0 keV X-ray luminosities of $3.1\times 10^{39}$ erg s$^{-1}$  for the ISM 
in NGC~1400  and  $4.3\times10^{39}$ erg s$^{-1}$ for the enhanced region.  
Using the {\sl WISE} image, we determined the $K$-band luminosity of NGC~1400 to be 
1.1$\times10^{10}$ $L_{K_\odot}$ within two optical  effective radii. 
We compared the  X-ray-infrared luminosity ratio ($L_X/L_K$) of NGC~1400 to
those of other early-type galaxies studied by Su \& Irwin (2013).  
There is a large scatter in $L_X/L_K$ and NGC~1400 is near the bottom of the range. 
After adding the X-ray luminosity of the enhanced region to that of NGC~1400's ISM, 
the $L_X/L_K$ ratio of NGC~1400 becomes closer to the average value of these systems. 
This supports the possibility that the enhanced region consists of 
ISM stripped from NGC~1400. 

Given that its X-ray emission is extended, this enhanced region may be a background galaxy cluster or group. 
If so, it should host at least one galaxy as luminous as NGC~1407, which has an 
absolute magnitude of $M_K$= -25.74 within two effective radii (Su \& Irwin 2013;  $r_{\rm e}=1.17^{\prime}$, from the RC3). 
Inspection of archived {\sl Two Micron All Sky Survey} ({\sl 2MASS})  images
revealed no infrared counterpart anywhere within the enhanced region. 
Since {\sl 2MASS} has a detection limit of $m_K$=14.3, this implies that a putative background galaxy 
should be further than $D_L\gtrsim1000$ Mpc to be undetected. 
At this distance, the enhanced region would have a physical radius of 300 kpc, 
too small for typical clusters but comparable to some small groups.  
To test whether the enhanced region could be a galaxy group at this distance,
we fit its X-ray spectrum to an {\tt apec} thermal emission model with 
a redshift of 0.24 ($D_L\approx1000$ Mpc). 
We obtained a best-fit $L_{\rm X, bol}$ of $\sim1.0\times10^{43}$ erg s$^{-1}$ and  temperature of 1.4 keV, typical of a galaxy group. But the best-fit metal abundance is 1.6$^{+1.1}_{-0.6}$ Z$_{\odot}$, too large for the metallicity of intragroup medium (Sasaki et al.\ 2013).
We also allowed the redshift in the {\tt apec} thermal emission model to freely vary and obtained a best-fit redshift of -0.008, temperature of 0.81 keV and metallicity of 0.48 Z$_{\odot}$. 
Consequently, we believe that the enhanced region is unlikely to be a background object. 

\begin{figure}
\epsscale{1.1}
\plottwo{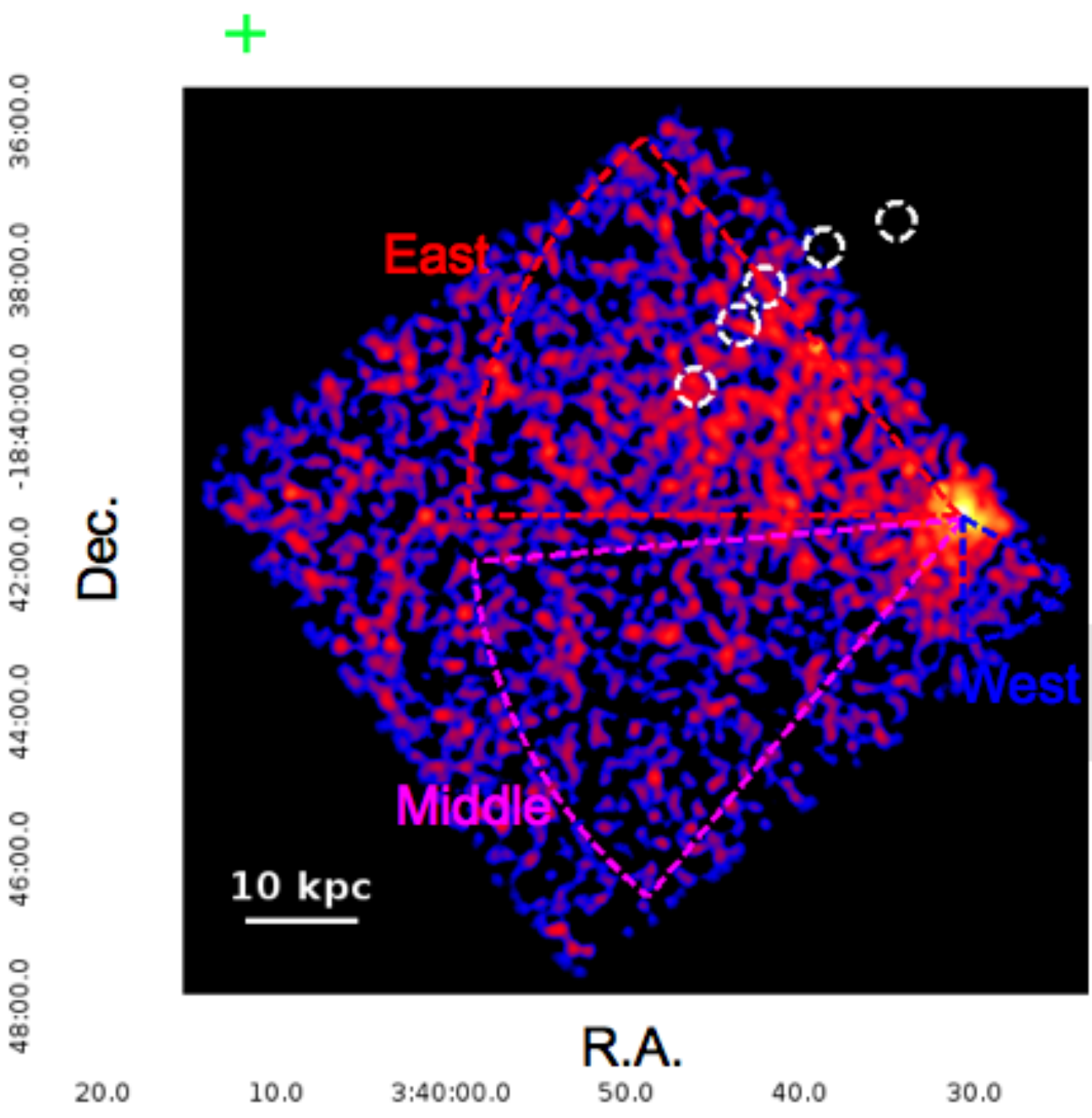}{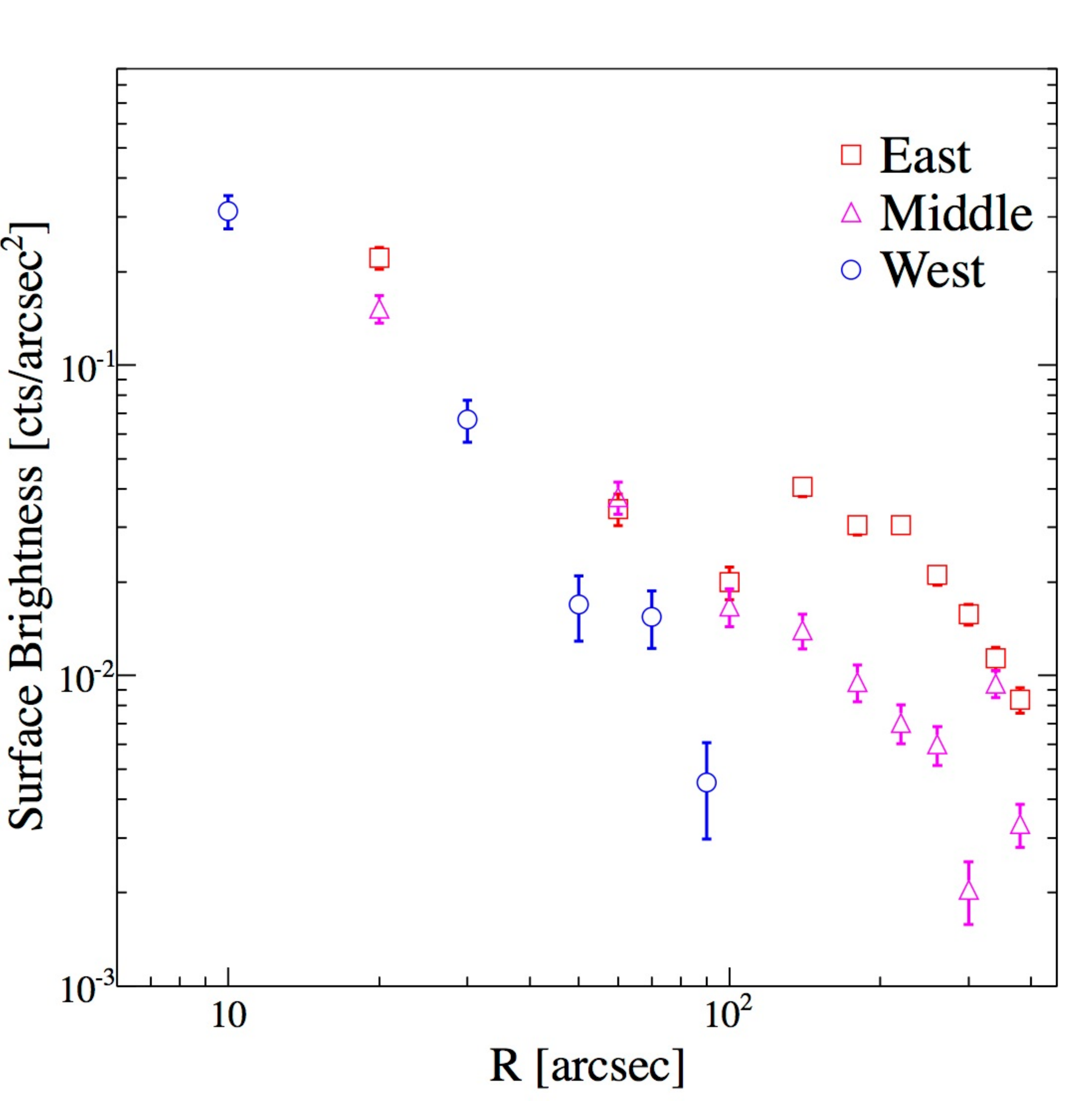}
\caption{\label{fig:sur} {\sl left}:  {\sl Chandra} image ($0.5-2.0$ keV) of NGC~1400 and the region of enhanced surface brightness. 
Green cross in the upper left corner: position of NGC~1407. White circles: positions of ``knots" found in the {\sl XMM-Newton} images; most of them are resolved to be point sources with the {\sl Chandra} image.  
{\sl right}: Surface brightness profiles centered on NGC~1400, averaged over the sectors labeled east, west and middle on the {\sl Chandra} image. [{\sl see the electronic edition of the journal for a color version of this figure.}]} 
\end{figure}

The {\sl XMM-Newton} image suggests that there is a sharp edge on the eastern side of the enhanced region. 
We found from {\it Chandra} imaging that  the several knots along this apparent edge in 
the {\it XMM-Newton} image are actually point sources. 
Figure~\ref{fig:sur} shows the {\it Chandra} ACIS-S image in the 0.5--2.0 keV band, after point sources were removed. 
Figure~\ref{fig:sur} also shows surface brightness profiles centered on NGC~1400, extending across the enhanced region, as well as toward other directions.
With point sources removed,
there is no sharp edge on the eastern side of the enhanced region. 

\subsection{\sl Thermal structures in the group gas}

\begin{figure} 
\epsscale{1.0}
\plotone{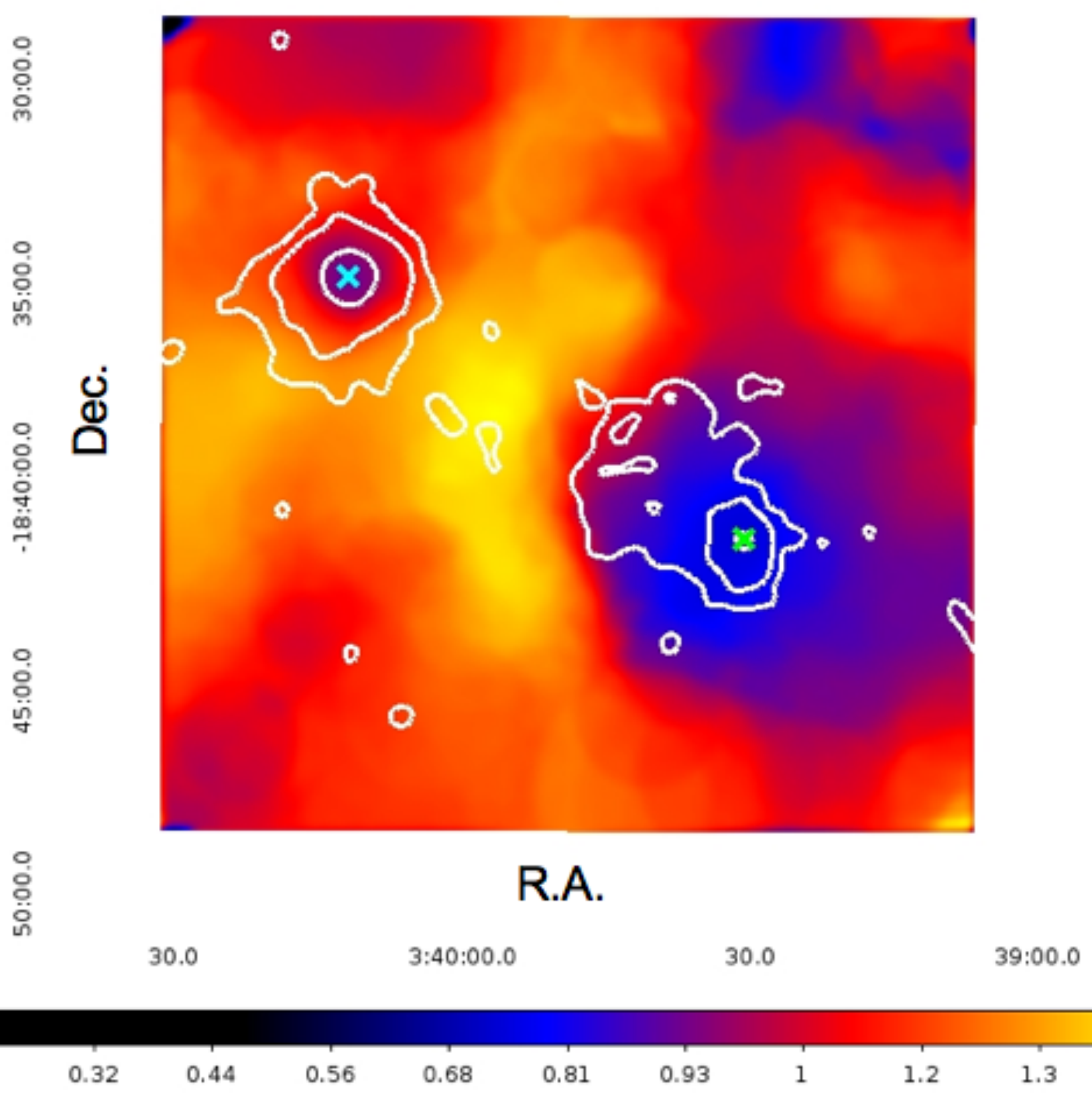}
\caption{\label{fig:t1} Temperature map (in keV) for the NGC~1407/1400 complex obtained with the western 
{\sl XMM-Newton} pointing using adaptive binning.  
[{\sl see the electronic edition of the journal for a color version of this figure.}]}
\end{figure}

In \S2.3.1 we described  the creation of an adaptively-binned temperature map 
(see Figure~\ref{fig:t1}) of the 
NGC~1407/1400 complex, using the western {\sl XMM-Newton} pointing. 
In this map the average temperature of the group gas is $\sim1$ keV and
there is an obvious cool core associated with NGC~1407. 
NGC~1400 and the enhanced region have a lower temperature 
($\sim0.7-0.8$ keV) than the group gas. 
There appears to be an arc of hotter gas ($\sim1.3$ keV) ranging from
southeast to north of NGC~1400.  
There are several possibilities for the presence of this hot arc. 
First, this is a cool core system; such systems generally have 
temperature profiles  rising from the center out to 0.1--0.3 virial radii, 
then declining outward, tracing the gravitational well 
(see Vikhlinin et al.\ 2006 for cool-core clusters and Sun et al.\ 2009 for
cool-core groups). 
In this scenario, this hot arc corresponds to the annulus of the maximum temperature.  
A second possibility is that NGC~1400 is moving eastward, as well as along the line-of-sight, and colliding with the NGC~1407 group atmosphere, consequently  causing such a hot arc. 
A third possibilty is that NGC~1400 has already passed through the inner atmosphere of NGC~1407 and is moving westward as well as along the line-of-sight, leaving a heated region behind;
this would be analogous to the motion of NGC~541 in Abell~194 (Bogdan et al.\ 2001).

\begin{figure} 
\epsscale{0.9}
\plotone{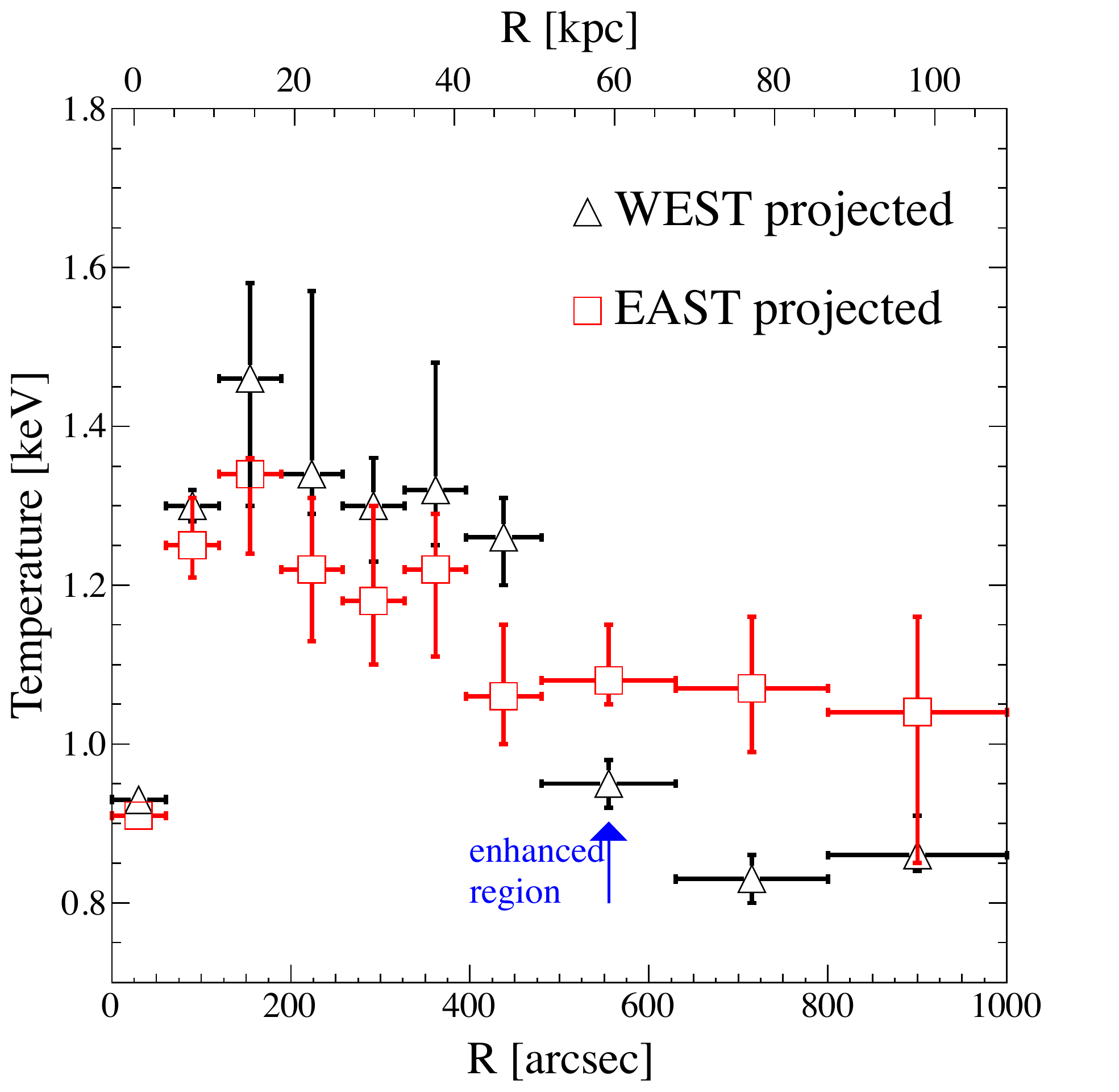}
\caption{\label{fig:tp} Projected temperature profiles centered on NGC~1407 and extending eastward and westward.
[{\sl see the electronic edition of the journal for a color version of this figure.}]}
\end{figure}

To investigate the origin of this hot arc, we performed a more azimuthally complete study of the thermal structure of the NGC~1407 group by combining the {\sl XMM-Newton} observations on both sides of NGC~1407. 
We first derived the projected temperature profiles for the eastern and western sides, as shown in Figure~\ref{fig:tp}. 
The temperature profiles within 7$^{\prime}$ (46 kpc) of NGC~1407 on both sides  are consistent within their uncertainties. 
The hot arc between the enhanced region and NGC~1407 found in Figure~\ref{fig:t1} corresponds to western bins 3--6, as shown in Figures~\ref{fig:14072} and \ref{fig:tp}. 
These regions are somewhat warmer, but consistent within the uncertainties, compared to the other side of NGC~1407 at the same radii. 
Furthermore, as we discuss later, shock heating in this system may not be observable, due to projection effects.  
Nonetheless, the 7th radial bin ($r=7.3^{\prime}=46$ kpc;  just east of the enhanced region) in the western pointing 
is hotter than the annular section at the same radius in the eastern pointing. 
Although there is no {\it a priori} reason to expect NGC~1407 group to have a perfectly symmetric temperature profile, 
we discuss possible causes of this localized heating in the Discussion section.

\subsection{\sl Hydrostatic mass estimate for the NGC 1407 group}

We derived the deprojected temperature and electron density profiles of the NGC~1407 group from a deprojected spectral analysis of the eastern {\sl XMM-Newton} pointing, as shown in Figures~\ref{fig:t} and \ref{fig:ne}.
We fit the deprojected temperature profile with the three-dimensional temperature profile calibrated by Vikhlinin (et al.\ 2006): 
\begin{equation}
T(r)=\frac{[T_0(r/r_{\rm cool})^{a_{\rm cool}}+T_{\rm min}]}{[(r/r_{\rm cool})^{a_{\rm cool}}+1]}\frac{(r/r_t)^{-a}}{[1+(r/r_t)^b]^{c/b}};
\end{equation}
we obtained a best-fit of [$T_0, r_{\rm cool}, a_{\rm cool}, T_{\rm min}, r_t, a, b, c$] = [$2.68, 81.97, 0.024, 0.180, 167.1, -0.34, 10, 0.65$].
We fit the deprojected electron density profile to a single $\beta$-model in the form of  
\begin{equation}
n_{\rm e}(r)=n_{\rm e0}[1+(r/{r_0})^2]^{-3\beta/2},
\end{equation}
obtaining best-fit values  of $n_{\rm e0}=0.96~\rm cm^{-3}$, $r_0$=0.174  kpc, and $\beta$=0.47.

\begin{figure} 
\epsscale{1.0}
\plotone{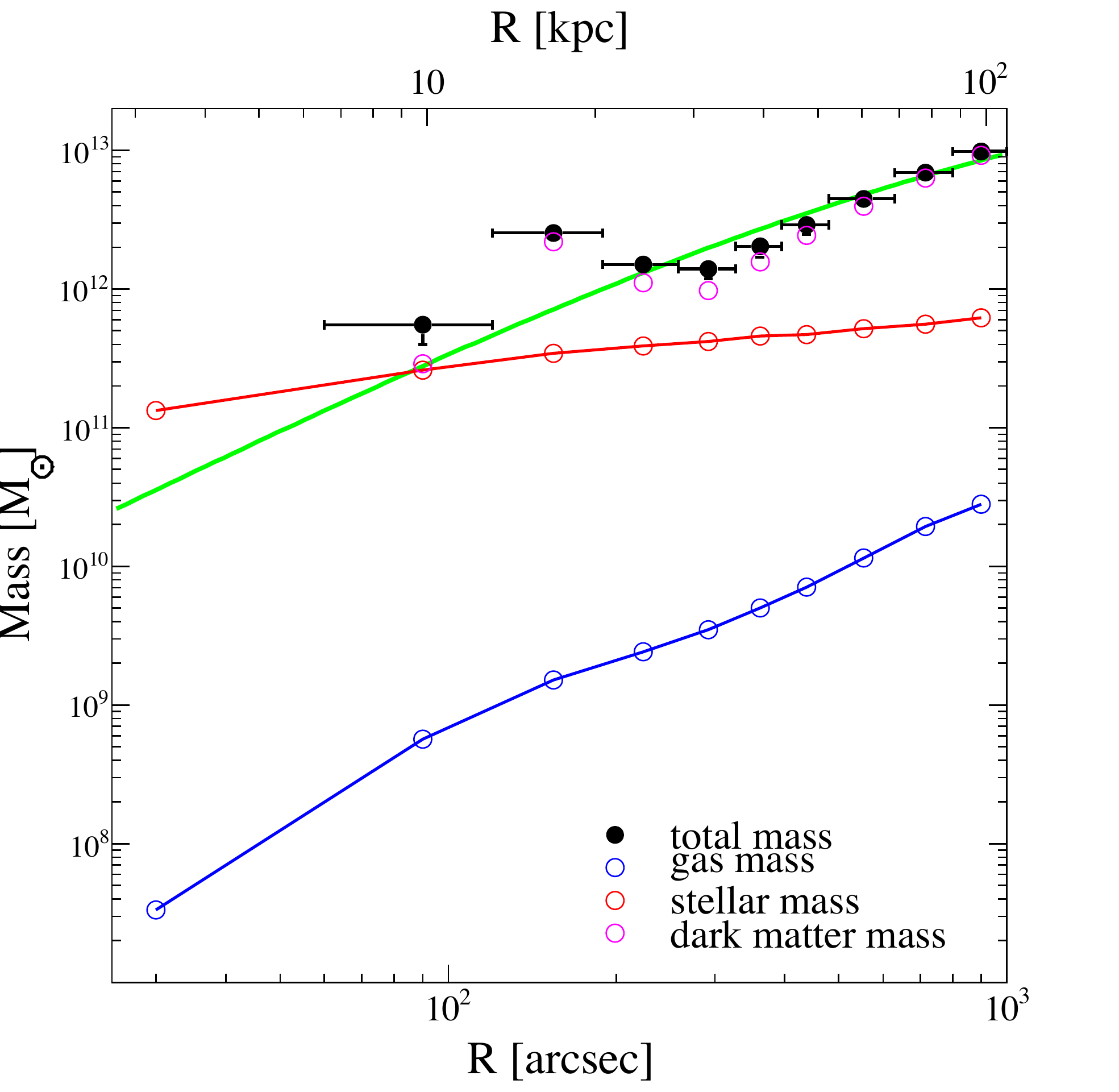}
\caption{\label{fig:mass} Black solid circles: profile of the enclosed total mass of the NGC~1407 group, 
as derived from the eastern {\sl XMM-Newton} pointing. 
Red open circles: enclosed stellar mass profile. 
Blue open circles: enclosed gas mass profile.
Pink open circles: enclosed dark matter profile.
Green line: best-fit NFW dark matter mass profile, with a concentration of
 $c=12.11\pm 1.80$ and a scaling radius of $r_{\rm s}=56.2\pm13.8$ kpc.  
[{\sl see the electronic edition of the journal for a color version of this figure.}]}
\end{figure}

The deprojected temperature and density profiles derived from the more hydrodynamically relaxed eastern pointing 
were used to calculate the total mass distribution,
using the equation of hydrostatic equilibrium:
\begin{equation}
M(r)=-\frac{kT(r)r}{\mu m_p G}\left(\frac{{\rm d~ln~\rho_{g}(r)}}{{\rm d~ln}~r}+
\frac{{\rm d~ln}~T(r)}{{\rm d~ln}~r}\right). 
\end{equation}
Uncertainties in mass estimates were estimated through Monte Carlo realizations
of the temperature and density profiles. 
The total mass within 100 kpc is $1.0^{+0.06}_{-0.07}\times10^{13}M_\odot$ and the
associated enclosed baryon mass fraction is $f_b\sim0.06$. 
We found that varying the surface brightness of the X-ray background
components by 20\% affects our total mass estimate by $\sim10$\%. 
Total mass and gas/stellar mass profiles are shown in Figure~\ref{fig:mass}.
Our results are consistent, in the regions of overlap, with the joint {\sl ROSAT}/{\sl Chandra} 
studies of this group by Zhang et al.\ (2007).

We obtained the dark matter distribution in this group by subtracting the baryonic mass 
distribution from the total mass distribution. 
In Figure~\ref{fig:mass} we compare the enclosed dark matter mass profile to the  
quasi-universal enclosed dark matter mass profile of Navarro et al.\ (1997):
\begin{equation}
M(r)=4\pi  \delta_{c}\rho_c(z){r_s}^3m(r/r_s), ~~~~{\rm where}~~ m(x)={\rm ln}(1+x)-\frac{x}{1+x}.
\end{equation}
This is derived from the dark matter density profile 
\begin{equation}
 \rho(r)=\frac{\rho_{c}(z)\delta_{c}}{(r/r_{s})(1+r/r_{s})^{2}},
\end{equation} 
where 
$r_{\rm s}$ is a scaling radius, 
$ \rho_{c}(z)=3H(z)^{2}/8\pi G $,  and
$$ \delta_{c}=\frac{200}{3}\frac{c^3}{{\rm ln}(1+c)-c/(1+c)}, $$
from which we can determine the dark matter concentration $c$. 
We derived values of $c=12.11\pm 1.80$ and $r_{\rm s}=56.2\pm13.8$ kpc.  
The virial radius is $R_{\rm vir}=681\pm 193$ kpc, derived from the definition of 
$c$ ($\equiv R_{\rm vir}/r_s$);
the extrapolated 
total mass within $R_{\rm vir}$ is 3.75$\pm0.57\times10^{13}$ $M_\odot$.

We calculated the $K$-band mass-to-light ratio to be 15.8 $M_\odot/L_{K_\odot}$ 
within 100 kpc. 
This corresponds to a $B$-band $M/L_B$ of  $\sim80$ $M_\odot/L_{B_\odot}$, 
assuming $L_{K_\odot}/L_{B_\odot}=5$ 
(Nagino \& Matsushita 2009). 
Since we do not know the stellar mass of this group beyond 100 kpc,
the extrapolated total mass within the virial radius (3.75$\times10^{13} M_\odot$) 
provides an upper limit of 59.4 $M_\odot/L_{K_\odot}$ 
($\approx300$ $M_\odot/L_{B_\odot}$) for the total $M/L$  within $R_{\rm vir}$. 

\section{\bf Discussion}

\begin{figure} 

\epsscale{1}
\centering
\plotone{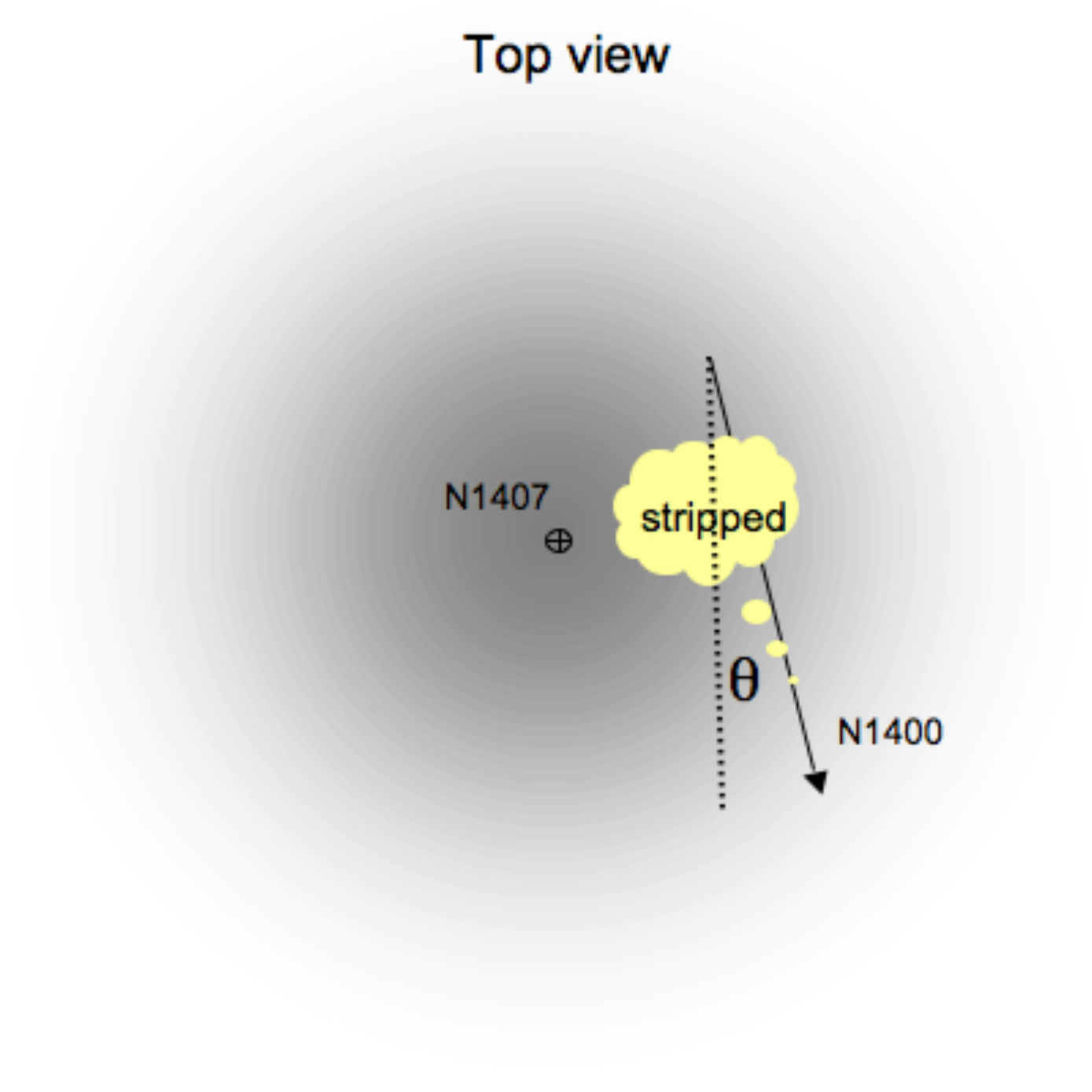}
\caption{\label{fig:tv} Illustration of NGC~1400 having its ISM partially stripped
out while moving through the group gas of NGC~1407 (top view, perpendicular to line of sight). 
[{\sl see the electronic edition of the journal for a color version of this figure.}]}
\end{figure}

Here we consider a scenario where NGC~1400 is moving towards us 
through the atmosphere of the 
NGC~1407 group, with a velocity mostly along the line-of-sight
and slightly westward, as illustrated in Figure~\ref{fig:tv}.
The outer ISM of NGC~1400 has been stripped and trails behind, 
creating the region of enhanced surface brightness east of NGC~1400.

\subsection{\sl The motion of NGC~1400 and the length of its stripped tail}

NGC~1400 has a line-of-sight velocity of 1200 km~s$^{-1}$ relative to NGC~1407. 
The projected distance between NGC~1407 and NGC~1400 is 78 kpc (11.9$^{\prime}$). 
Here we try to place some rough limits on the total velocity of NGC~1400 relative to NGC~1407.

The position of the enhanced region implies that NGC~1400 has a westward velocity component. 
We estimate the westward velocity of NGC~1400 through the
pressure jump it has induced beyond the western leading edge. 
We follow the treatment of Vikhlinin et al.\ (2001) for the motion
of a blunt body through intracluster gas: the pressure
difference between a distant ``free stream" region and the stagnation point
(where the local relative velocity is brought to zero at the body's leading edge)
is related to the velocity (Mach number) in that direction (Landau \& Lifshitz 1987):
\begin{equation}
\frac{p_1}{p_2}=\left(1+\frac{\gamma-1}{2}\mathcal{M}_{\rm ap}^{2}\right)^{\frac{\gamma}{\gamma-1}} 
~~~{\rm for}~ \mathcal{M}_{\rm ap} \le 1 ~{\rm (subsonic});
\end{equation}
\begin{equation}
\frac{p_1}{p_2}=\left(\frac{\gamma+1}{2}\right)^{\frac{\gamma+1}{\gamma-1}}\mathcal{M}_{\rm ap}^{2}\left(\gamma-\frac{\gamma-1}{2\mathcal{M}_{\rm ap}^2}\right)^{-\frac{1}{\gamma-1}}
~~~{\rm for}~ \mathcal{M}_{\rm ap} > 1 ~{\rm (supersonic)};
\end{equation}
where the adiabatic index $\gamma = 5/3$.
We were unfortunately unable to identify a clear leading edge,
likely due to the highly line-of-sight motion of NGC~1400 and 
the tenuousness of the group gas west of NGC~1400.
Instead, we obtained a constraint from the (factor of 3.4)  jump in surface brightness  in 
the {\sl Chandra} profile westward of NGC~1400 from 50$^{\prime\prime}$ to 100$^{\prime\prime}$,
which likely spans a range of radii from just within the stagnation radius to the free stream region.
We assume the density jump equals the square root of surface brightness jump
(thus, a factor of 1.85). 
The temperature of the group gas at the stagnation point ($\sim$1 keV) 
is calculated from eq.~(4) for regions at the same radius from NGC~1407 as NGC~1400. 
We take the temperature of the enhanced region (0.8 keV) as the 
temperature of regions just inside the leading edge. 
The apparent pressure jump is thus 1.5, corresponding to an apparent Mach number of 
$\mathcal{M}_{\rm ap}=0.74$. 
The sound speed of the NGC~1407 group gas is
$c_{\rm s}=({\gamma kT/\mu m_H})^{1/2}$ = 507 km~s$^{-1}$, 
where the group gas temperature $kT\approx1$ keV.
The inferred tangential velocity is $v_{\rm tan}$ = $\mathcal{M}_{\rm ap} c_{\rm s}$ = 
$0.74\times 507$~km\,s$^{-1}$ = 375~km\,s$^{-1}$. 
This gives us a total velocity of 1257~km\,s$^{-1}$ ($\mathcal{M}\approx2.5$) for NGC~1400, 
in a direction $\theta = 20^{\circ}$ from the line-of-sight.  
However, unlike cases in Abell~3667 (Vikhlinin et al.\ 2001), 
NGC~1404 (Machacek et al.\ 2005), and Abell~194 (Bogdan et al.\ 2011), 
the motion of NGC~1400 seems highly line-of-sight, which makes it difficult to 
infer its tangential velocity through this method.

We used the total mass profile of the NGC~1407 group (derived in \S3.3 
and indicated by the green line in Figure~\ref{fig:mass}) to provide
an additional constraint on the motion of NGC~1400. 
We first constrained the total velocity of NGC~1400 by assuming it resides at the same distance as NGC~1407, so their physical separation is the same as their projected distance. 
If NGC~1400 experienced free-fall from infinity to its current position, we deduced
from the group mass profile that NGC~1400's total relative velocity would be 1500 km\,s$^{-1}$. 
Considering that NGC~1400 has a line-of-sight velocity $v_{\rm los}=1200$ km~s$^{-1}$ and 
its projected separation from NGC~1407  is only a lower limit to NGC~1400's distance from the group center,
we infer that NGC~1400's velocity in the plane of sky is $<$ 900 km~s$^{-1}$ 
and $\vert \theta \vert < 36.9^{\circ}$. 
If we instead assume that NGC~1400 has no velocity component in the plane of sky 
($\theta$ $=$ 0$^{\circ}$), its current velocity ($v= 1200$ km~s$^{-1}$) then requires 
that NGC~1400 resides at a distance 400 kpc closer (or farther) than NGC~1407.
Given a possible velocity range of $1200-1500$ km s$^{-1}$, the 
Mach number $\mathcal{M}\equiv v/c_{\rm s}$ would be in the range of 2.4--3. 
Our previous estimate of the motion (Mach number) of NGC~1400, derived from the observed pressure jump
at its leading edge, lies in this range.

Since the strength of ram pressure stripping is proportional 
to the ambient gas density, significant stripping is most likely to occur
near the group center, given the low average gas density of this group.
Given the proximity of the stripped tail to NGC~1400,
this implies that NGC~1400 
likely resides near the group center, so its
projected distance from NGC~1407 should be comparable to its physical separation. 
This leads us to prefer the higher total velocity estimates. 

If the surface brightness enhancement east of NGC~1400 is indeed the stripped tail of NGC~1400 (see \S3.1), 
the direction of NGC~1400's motion allows us to infer the tail length, assuming the tail is aligned with the current direction of NGC~1400's motion. 
The projected length of NGC~1400's tail (the projected northeast-southwest extent of the enhanced region) is $\sim25$ kpc. 
Since NGC~1400 is estimated to be moving at an angle $\theta\approx30^{\circ}$ from the line-of-sight,
the actual tail length is $\sim50$ kpc.
In numerical simulations of ram pressure stripping, typical lengths of stripped tails are $\sim40$ kpc (Roediger \& Bruggen 2008). 
Observationally, there is a large scatter in reported tail lengths (Sun et al.\ 2007b) 
and some can be as long as $\sim380$ kpc (M86 -- Randall et al.\ 2008). 
There are various factors that can affect the length (or morphology) of stripped tails, such as the velocity of the moving galaxy, the time scale of the stripping process, the relative emissivities of the tail and the ambient ICM, and projection effects 
(Ruszkowski et al.\ 2012). 
In summary, our preferred scenario is illustrated 
in Figure~\ref{fig:tv}, where NGC~1400 is moving through the atmosphere of the
NGC~1407 group at an angle of $\sim30^{\circ}$ from the line-of-sight, 
with a total relative velocity of $\sim1400$ km~s$^{-1}$.

\subsection{\sl The apparent absence of shock heating}

Most merger shocks reported to date have been found in galaxy clusters (e.g.\ Abell~3376, Sarazin 2013), rather than in galaxy groups. 
Shocks found in groups are usually related instead to AGN feedback 
(HCG 62 -- Gitti et al.\ 2010). 
One exception is Abell~194, a poor cluster ($kT\sim2$ keV) observed to have a merger shock:
its member galaxy NGC~541 has a velocity of 788 km~s$^{-1}$ with respect to the cluster center and 
has a reverse shock with $\mathcal{M}\approx0.9$ (derived from a discontinuity in its pressure profile; Bogdan et al.\ 2011). 
In the NGC~1407 group, NGC~1400 has a velocity of at least 1200 km~s$^{-1}$ 
($\mathcal{M} \gtrsim 2.4$) relative to the group center. 
Thus, we expect to observe some evidence of a shock, such as density discontinuities and shock heating. 
From the Rankine-Hugoniot jump conditions (Laudau \& Lifshitz 1987), the expected density and temperature jumps can be expressed as a function of the Mach number: 
\begin{equation}
\frac{{\rho}_2}{{\rho}_1}=\frac{\mathcal{M}^2(\gamma+1)}{2+\mathcal{M}^2(\gamma-1)},
\end{equation}
\begin{equation}
\frac{T_2}{T_1}=\frac{[(\gamma-1)\mathcal{M}^2+2][2\gamma \mathcal{M}^2-(\gamma-1)]}{(\gamma+1)^2\mathcal{M}^2}.
\end{equation}
NGC~1400 is moving with $\mathcal{M}\gtrsim2.4$, so the expected density jump 
is $\rho_2/\rho_1\gtrsim 2.6$ and the 
expected temperature jump is $T_2/T_1\gtrsim2.6$.

Given that the group gas temperature is $\sim1$ keV, we expect to observe some gas heated to $\gtrsim2.6$ keV in the field of view.  
However, we do not detect obvious shock features. 
A similar situation has been reported for M86 in the Virgo cluster. 
It has a line-of-sight velocity difference of 1550 km~s$^{-1}$ 
with respect to M87 at the cluster center, which is almost twice the sound speed of the cluster gas (850 km~s$^{-1}$). 
In spite of a Mach number $\mathcal{M}\approx2$, no shock feature has been found 
around M86 (Randall et al.\ 2008);
the absence of shock features has been attributed to projection effects (Mazzotta et al.\ 2001; Rangarajan et al.\ 1995). 
Akahori \& Yoshikawa (2010) 
simulated cluster shock features as a function of viewing angle.
They found that sharp surface brightness discontinuities at shock layers are clearly visible  only when the collisional direction is nearly perpendicular to the line-of-sight 
($\theta$ = 90$^{\circ}$). 
The apparent Mach number 
can be reduced by more than 60\% when 
$\theta$ is smaller than 30$^{\circ}$. 
The motions of M86 and NGC~1400 are nearly line-of-sight ($\theta$ $\lesssim30^{\circ}$), 
so shock features may be smeared out by projection effects. 
This is in sharp contrast to the case of Abell~194, where the motion of NGC~541 is 
almost perpendicular to the line of sight (Bogdan et al.\ 2011).
Our current X-ray data do not allow us to constrain spectrally whether there is a layer of 
shocked group gas superposed on ISM stripped from NGC~1400, as well as unshocked group gas.

\begin{figure}
\epsscale{1.1}
\plottwo{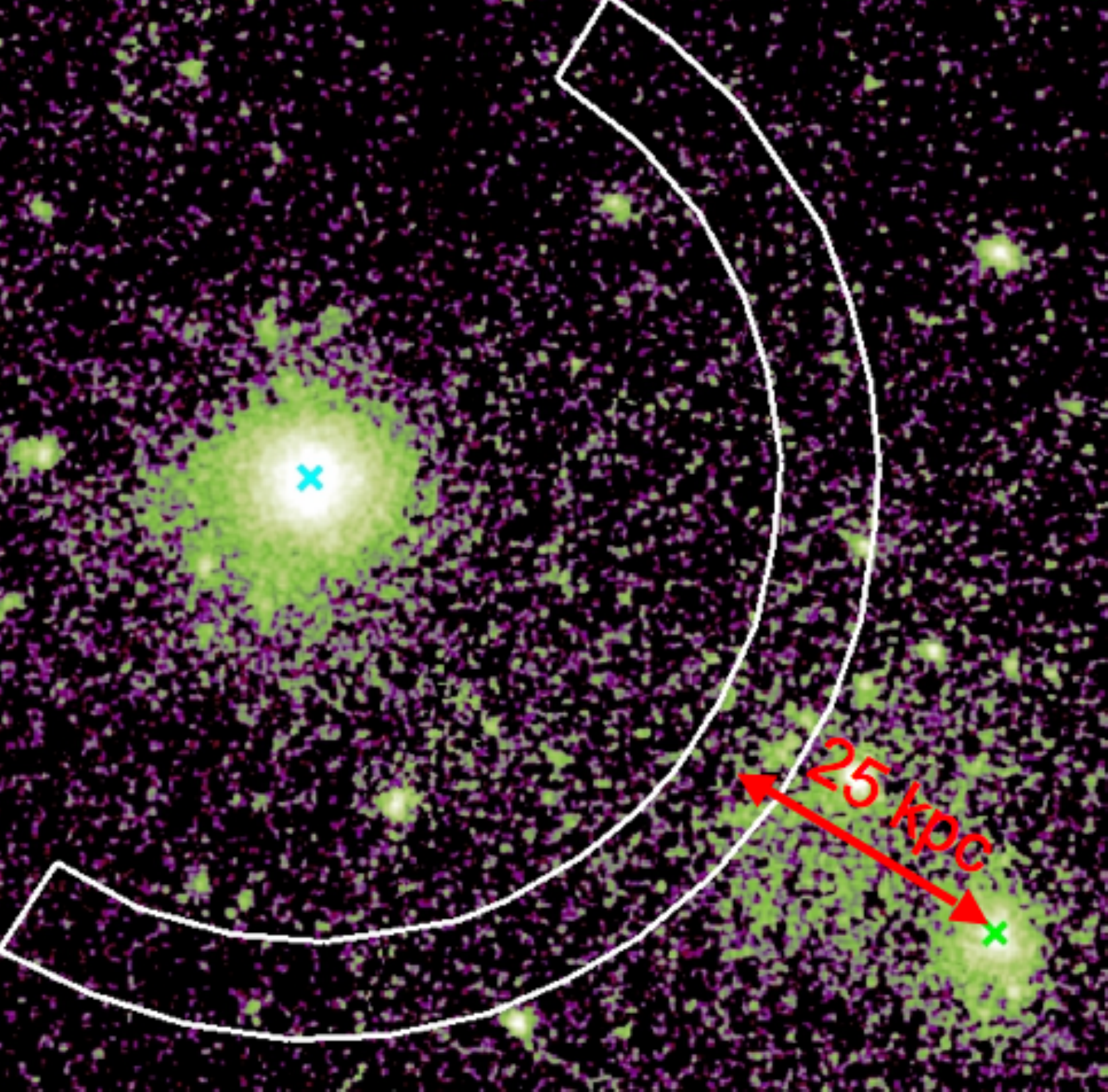}{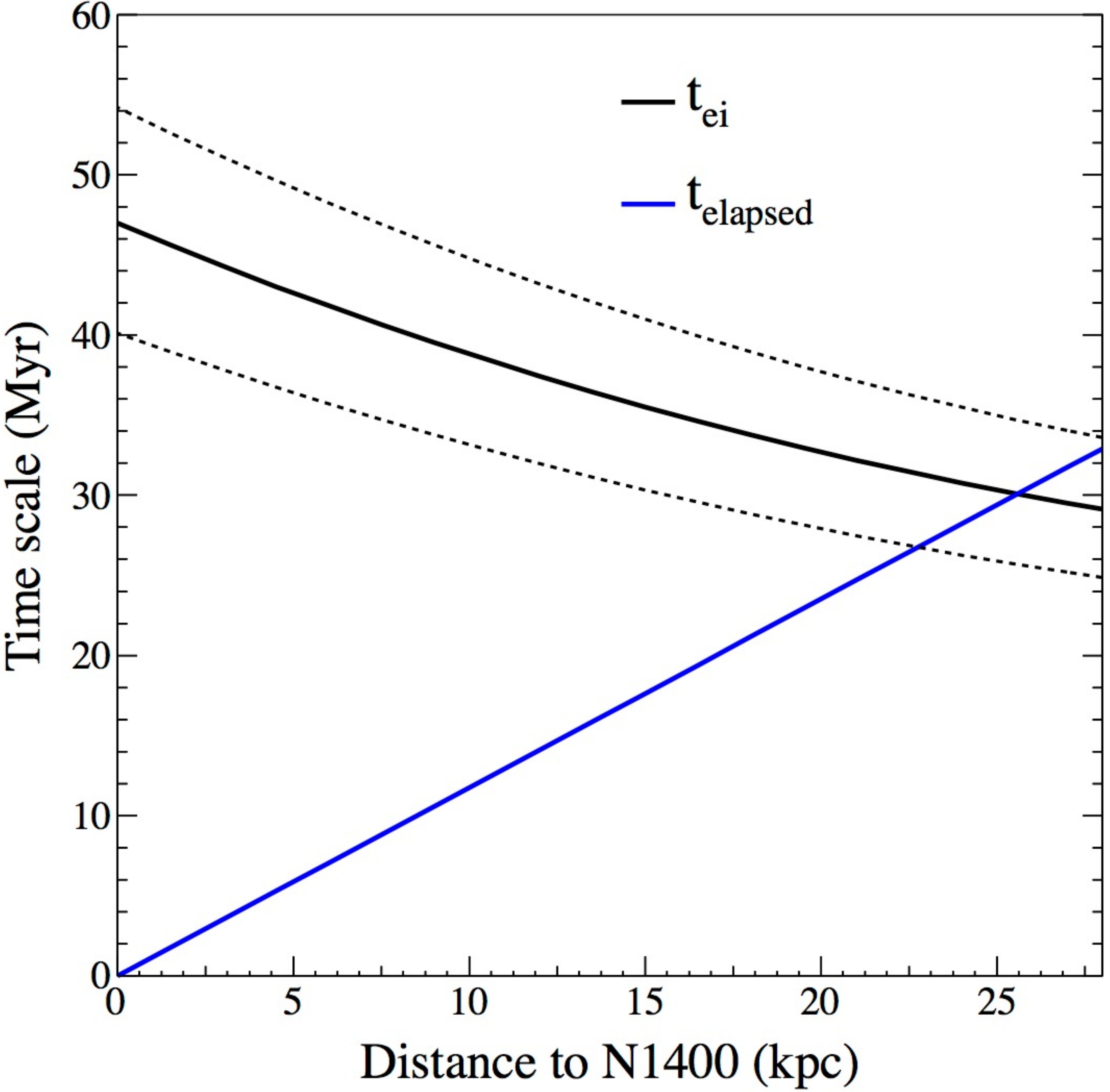}
\caption{\label{fig:ts} 
$left$: {\sl XMM-Newton} image of NGC~1407/1400 (0.5-2.0 keV band). 
White annular section: the 7th bin of western pointing,  which exhibits heating. 
Red line: Proposed path of ISM stripped from NGC~1400.
Cyan cross: position of NGC~1407. 
Green cross: position of NGC~1400.  
{\sl right}: $t_{\rm ei}$ and $t_{\rm elapsed}$ as functions of distance from NGC~1400 
across the enhanced region, as shown in the red line in the {\sl left} image. 
The 7th bin (beyond 25 kpc) is most likely to be   in thermal equilibrium. 
[{\sl see the electronic edition of the journal for a color version of this figure.}]} 
\end{figure}

In addition to being washed out by projection effects, the observability of
a shock may be affected by  ions and electrons being out of thermal equilibrium.
The ion temperature may indicate a shock, but it takes a finite time
for the electrons to equilibrate with the ions and also exhibit the shock temperature.
Most of the shock energy would be initially 
converted to the thermal energy of ions in the post-shock region, due to the mass difference between ions and electrons 
(Spitzer 1962; Wong \& Sarazin 2009; Akahori \& Yoshikawa 2010).  
The time scale for electron-ion thermal equilibration through Coulomb collisions is given by
\begin{equation}
t_{\rm ei} \approx6.3 \times 10^7  \frac{(T_{\rm e}/10^7 ~{\rm K})^{3/2}}{(n_{\rm e}/10^{-4}~{\rm cm}^{-3})({\rm ln}~\Lambda/40)}~{\rm yr},
\end{equation}
(Rudd \& Nagai 2009), where ln~$\Lambda$ $\approx40$ is the Coulomb logarithm. 
Electron-ion thermal equilibrium usually is achieved in the inner regions of galaxy clusters, where the equilibration time is shorter than the dynamical time scale. 
Departure from this equilibrium has been proposed to occur in the outskirts of galaxy clusters, 
where the electron density is small and the equilibration time is relatively long 
(Fox \& Loeb 1997; Hoshino et al.\ 2010). 
The group gas of NGC~1407 has a smaller gas density ($n_{\rm e}\sim10^{-4}$ cm$^{-3}$) 
compared to similar regions in galaxy clusters. 
Thus, we examined the thermal equilibrium condition for this system. 
We assume that the shock has started at what is now the end of the tail 
(corresponding to the 7th bin of the western pointing) and propagate following the motion of NGC~1400, as indicated in Figure~\ref{fig:ts} ({\sl left}). 
For regions between the end of the tail  and NGC~1400,  
we calculated $t_{\rm ei}$ as a function of (projected) distance to 
NGC~1400\footnote{$n_{\rm e}$ and $T_{\rm e}$ are calculated from eq.~(4) and eq.~(5) respectively, derived with the eastern pointing}. 
We estimated the time  elapsed since the passage of a putative shock  ($t_{\rm elapsed}$),
assuming NGC~1400 has tangential velocity of 800 km~s$^{-1}$ 
along the (projected) length of the stripped tail. 
We compare $t_{\rm ei}$ and $t_{\rm elapsed}$ in Figure~\ref{fig:ts} ({\sl right}). 
For regions within  25 kpc (projected) of NGC~1400, the electrons might not have had enough time to equilibrate with ions, consequently leading to the absence of evidence for shock heating. 
The end of the tail (beyond 25 kpc) should be in thermal equilibrium. 
Interestingly, we did observe slight extra heating (temperature increasing by 0.2 keV or $\sim20$\%) in the 7th bins with {\sl XMM-Newton} observations. 
In principle, we should observe transitional shock heating regions 
along the path of NGC~1400 (on its eastern side), as ion-electron equilibrium is gradually established. 
Unfortunately, these {\sl XMM-Newton} and {\sl Chandra} data do not allow us to resolve this;
we would need deeper {\sl Chandra} observations with more areal coverage to test this.

\subsection{\sl Stripping conditions}

\subsubsection{\sl Ram pressure stripping}

For a consistency check, we studied the conditions for ram pressure stripping in the NGC 1407 group. 
The ISM of NGC~1400 would be stripped when the ram pressure ($P_{\rm ram}=\rho_{\rm gas}v^2$) exceeds the gravitational restoring force per unit area (Gunn \& Gott 1972): 
\begin{equation}
P_{\rm ram} > \frac{F}{A} ~\Rightarrow ~\rho_{_{\rm ICM}}v^2 > \frac{GM_{\rm tot}}{{R_{_{\rm ISM}}}^2}\frac{M_{_{\rm ISM}}}{\pi {R_{_{\rm ISM}}}^2},
\end{equation}
for a galaxy with a total mass of $M_{\rm tot}$ and 
a characteristic radius of 
$R_{_{\rm ISM}}$ (the radius of the galaxy at which the stripping occurs),
moving with total velocity $v$ through cluster gas with density $\rho_{_{\rm ISM}}$.
While this was originally derived for disk galaxies,
McCarthy et al.\ (2008) developed an analogous model for the ram pressure stripping of galaxies with spherically-symmetric gas distributions. 
Their model, which is more suitable for early-type galaxies with an extended atmospheres, 
yields the ram pressure stripping condition:

\begin{equation}
P_{\rm ram} = \rho_{_{\rm ICM}}v^2 >\frac{\pi}{2}\frac{GM_{\rm tot}\rho_{_{\rm ISM}}}{R_{_{\rm ISM}}},
\end{equation}
where 
$\rho_{_{\rm ISM}}$ is the   ISM gas density in the galaxy.

To examine this condition, we first fit the surface brightness profile of the ``middle" direction from NGC~1400 (shown in Figure~\ref{fig:sur}) to a $\beta$-profile: 
\begin{equation}
I(r)=I_{0}\left[1+\left(\frac{r}{r_c}\right)^2\right]^{-3\beta/2+1/2}.
\end{equation}
We obtained $\beta$=0.27 and $r_c$=0.93 kpc.
Assuming isothermal gas, this corresponds to an ISM density profile of 
\begin{equation}
\rho_{_{\rm ISM}}(r)=\rho_{0}\left[1+\left(\frac{r}{r_c}\right)^2\right]^{-3\beta/2}.
\end{equation}

The X-ray luminosity of the enhanced region is comparable to the hot gas luminosity of NGC~1400. 
Thus, roughly half of NGC~1400's ISM has been stripped, corresponding to a radius ranging from 6.4 kpc to 8.7 kpc, derived from the volume integration of the ISM density profile. 
Thus, we adopted a characteristic radius of $R_{_{\rm ISM}}$=7.6 kpc for this ram pressure stripping process. 
This calculation also lead us to obtain the value of $\rho_{0}$=0.012 cm$^{-3}$, using the ISM mass of NGC~1400, determined through the best-fit normalization in the spectral analysis of \S3.1. 
The density of group gas is chosen to be $\rho_{_{\rm ICM}}=2.5\times 10^{-4}$ cm$^{-3}$, calculated from eq.~(5) for regions at the same radius from NGC~1407 as the enhanced region. 
The ISM density $\rho_{_{\rm ISM}}=2.2\times 10^{-3}$ cm$^{-3}$ is calculated from eq.~(21) at a radius of $R_{_{\rm ISM}}$=7.6 kpc. 
We estimated the total mass of NGC~1400 enclosed within a spherical radius $r$ through the hydrostatic equilibrium 
equation shown in eq.~(6).
We chose $r$ to be the same as $R_{_{\rm ISM}}$=7.6 kpc. The gas density gradient is chosen to be -3$\beta = -0.81$ at this radius
and we assume the temperature gradient is is zero. 
We obtained a total enclosed mass of $M_{\rm tot}$ = 1.5$\times 10^{11}$ $M_\odot$.
We calculated through eq.~(19) that the relative velocity of NGC~1400 needs to be $\sim1275$ km\,s$^{-1}$ to have the gas in the enhanced region stripped via ram pressure.  This value lies in a possible range (1200 -- 1500 km\,s$^{-1}$) of its total velocity estimated in \S4.1. 
Therefore, ram pressure stripping alone is capable of forming the enhanced region.

Once the stripping condition is satisfied, we can estimate the time scale for 
the ram pressure stripping process, which is given by
\begin{equation}
t_{\rm ram}=({\rm d}~{\rm{ln}}~m_{_{\rm ISM}}/{\rm d}t)^{\rm -1},
\end{equation}
so that 
\begin{equation}
t_{\rm ram}\approx~\frac{R}{v}\left(\frac{2\rho_{_{\rm ISM}}}{\rho_{_{\rm ICM}}}\right)^{1/2} \approx3\times10^7  \left(\frac{\rho_{_{\rm ISM}}}{\rho_{_{\rm ICM}}}\right)^{1/2}\left(\frac{\it v}{10^3~ \rm km~s^{-1}}\right)^{-1} \left(\frac{\it R}{20~ \rm kpc}\right) {\rm yr}.
\end{equation}
(Takeda et al.\ 1984). It takes $\sim30$ Myr to have the hot gas in the enhanced region stripped from NGC~1400 through ram pressure.
Given that NGC~1400 may have a tangential velocity of $\approx800$ km~s$^{-1}$ (\S4.1), it may have travelled for $\gtrsim25$ kpc in the plane of sky within this time. 
This is in excellent agreement with the projected tail length.

\subsubsection{\sl Turbulent and Viscous Stripping}

Ram pressure is not the only possible stripping mechanism. 
There are other processes such as turbulence and viscosity   
which may contribute to the stripping of ISM, as noted by Nulsen (1982). 
The Reynolds number ($Re$) is the ratio of inertial to viscous forces, which can be obtained through 
\begin{equation}
Re = 2.8\,(r/\lambda)(v_{\rm gal}/c_{\rm s}) 
\end{equation}
(Batchelor 1967); $r$ and $v_{\rm gal}$ are the radius and velocity of the moving galaxy respectively; $\lambda$ is the effective mean free path of ions in the hot gas, which can be calculated through 
\begin{equation}
\lambda = 23 \left(\frac{T}{10^8\,{\rm K}}\right)\left(\frac{n_{\rm e}}{10^{-3}\,{\rm cm^{-3}}}\right)^{-1}\,\rm kpc 
\end{equation}
Sarazin (1988).

For low Reynolds numbers ($\sim$ 10), laminar viscous flow occurs, characterized by smooth constant fluid motion, while for high Reynolds numbers ($\sim$ 2000), turbulent flow occurs, producing various instabilities. 
For NGC~1400 we assume $r=8.6$ kpc and $v=1400$ km\,s$^{-1}$; 
$T$ and $n_{\rm e}$ are taken to be 1.1 keV and 2.5$\times 10^{-4}$\,cm$^{-3}$, respectively, derived from eqs.\ (4) and (5);
thus, we estimate that $Re = 5.9$, so laminar viscous stripping is preferred 
(in the absence of magnetic fields).

For turbulent stripping, the typical  mass-loss rate of the ISM is approximately (Nulsen 1982):
\begin{equation}
\dot{M}_{\rm tur}\approx\pi r^2\rho_{_{\rm ICM}} v_{\rm gal} =0.69\left(\frac{n_{e,_{\rm ICM}}}{10^{-3}\,\rm cm^{-3}}\right)\left(\frac{r}{2.5\,\rm kpc}\right)^2\left(\frac{v_{\rm gal}}{1200\,\rm km\,s^{-1}}\right) M_\odot\,{\rm yr^{-1}}.
\end{equation}
This gives a typical stripping time scale of (Sun et al.\ 2007b):
\begin{equation}
t_{\rm strip}=\int{\frac{d M}{\dot{M}}}=\frac{4}{n_{_{\rm ICM}}v_{\rm gal}}\int n(r)dr$$
$$=0.224\,g_1\left(\frac{n_{\rm e0}}{0.2~ \rm cm^{-3}}\right)\left(\frac{n_{_{\rm ICM}}}{10^{-3}\,\rm cm^{-3}}\right)^{-1}\left(\frac{r_0}{0.4\,\rm kpc}\right)\left(\frac{v_{\rm gal}}{1400~ \rm km\,s^{-1}}\right)^{-1} \rm Gyr,
\end{equation}
$${\rm where} ~~g_1=\int(1+x^2)^{-1.5\beta}dx, ~x=r/r_0,$$
and $r_0$ and $\beta$ are the parameters in the $\beta$-model of the deprojected electron density profile. 
A stripped radial range from 6.4 to 8.7 kpc corresponds to $g_1=0.41$.  
We estimate from eq.~(27) that the enhanced region has been stripped during the last $19$ Myr.

For  viscous stripping, the typical mass-loss rate of the ISM is approximately (Nulsen 1982):
\begin{equation}
\dot{M}_{\rm vis}\approx\pi r^2\rho_{_{\rm ICM}} v_{\rm gal}\,(12/Re)
\end{equation}
In the case of NGC~1400, $\dot{M}_{\rm vis} \approx 2.1 \dot{M}_{\rm tur}$, 
so the time scale for viscous stripping is half that of turbulent stripping.

Although analytical methods show that viscous stripping can be very efficient in 
extracting gas from galaxies (e.g., Nulsen 1982), some (hydro)dynamical simulations 
indicate that magnetic fields could greatly reduce the effects of viscosity
(e.g., Roediger \& Bruggen 2008). 
Measurements of radio continuum emission and Faraday rotation show that magnetic fields 
commonly exist in galaxy clusters, with strengths of the order of 1$\mu$G 
(Govoni \& Feretti 2004; Bonafede et al.\ 2010). 
To include the effects of magnetic fields, we invoke a suppression factor 
($f_{\rm v}$), relating a corrected Reynolds number to the original unmagnetized version: 
$Re_{\rm m}=Re\,f_{\rm v}^{-1}$. 
Narayan \& Medvedev (2001) found $f_{\rm v}$ of 0.01--0.2 in intracluster gas,
so the corrected Reynolds number $Re_{\rm m}$ should be in the range of 30--550 
for NGC~1400 and the enhanced region. 
This points to a more turbulent, less viscous case compared with our original estimate. 
Roediger \& Bruggen (2008) demonstrate through simulations that intracluster gas flows 
through galaxies more smoothly in the viscous case, 
while Kelvin-Helmhotz and Rayleigh-Taylor instabilities occur when the viscosity 
is suppressed, leading to vortices and turbulence.

\subsection{\sl $M/L$ and the group membership of NGC~1400}

Romanowsky et al.\ (2009) studied this group using the kinematics of its globular clusters within 60 kpc. 
They extrapolated a virial mass of $\sim6\times10^{13}$ $M_\odot$ and an associated $M/L$ of  $\sim800$ $M_\odot/L_{B_\odot}$. 
This large $M/L$ would make NGC~1407 an unusually dark system and may require a large amount 
of baryons  to be lurking in an undetected phase (Romanowsky et al.\ 2009). In our study, we derived the hydrostatic mass of this group with an {\sl XMM-Newton} pointing east of NGC~1407. We determined a total mass within 100 kpc of $\sim1\times10^{13} M_\odot$ and an extrapolated virial mass of only $3.75\times10^{13}$ $M_\odot$ for this group. 
The disagreement between the total mass estimated through globular cluster kinematics and that 
determined through X-ray analysis of hot gas  is rather common for early-type galaxies,
some of which are at group centers 
(e.g, M87: Murphy et al.\ 2011; NGC~4636: Johnson et al.\ 2009; NGC~4649: Shen \& Gebhardt 2010; 
NGC~3923: Norris et al.\ 2012). 
The biggest problem with the X-ray modeling involves the possible lack of hydrodynamic equilibrium 
or the presence of non-thermal pressure support, due to magnetic fields, gaseous turbulence or cosmic rays (Churazov et al.\ 2010; Shen \& Gebhardt et al.\ 2010). 
Some studies show that the contribution from non-thermal pressure is much smaller than thermal pressure for early-type galaxies (Brighenti et al.\ 2009), as well as within $R_{500}$ for groups and clusters (Shaw et al.\ 2010).  Moreover, the eastern side of NGC~1407 that we used to derive a hydrostatic mass appears relatively relaxed, thus is likely to be close to hydrostatic equilibrium.  
On the other hand, the biggest concerns with the globular cluster probe include the unknown galaxy inclination and potentially complex orbit structure (Gavazzi 2005; Thomas et al.\ 2007). 
Romanowsky et al.\ (2009) found that their results are in better agreement with the previous
 {\sl Chandra} X-ray study (Zhang et al.\ 2007, within 20 kpc of NGC~1407) 
 if  the globular clusters are assumed to have a peculiar orbit distribution. 
In short, we believe that the total mass estimated through X-ray analysis may be relatively more robust in this case. 

Nevertheless, even a virial mass as low as $3.75\times10^{13}$ $M_\odot$ may still be able to keep NGC~1400 bound. 
According to Romanowsky et al.\ (2009), the minimum virial mass necessary to keep NGC~1400 bound is only $3\times10^{13}$ $M_\odot$, provided that the apocenter of NGC~1400 is at $R_{\rm vir}$. 
The upper limit of $M/L\approx300$ $M_\odot/L_{B_\odot}$ that we determined with the hot gaseous X-ray emission is 
half that estimated by Romanowsky et al.\ (2009), although their uncertainty is as large as a factor of two. 
Our results suggest that NGC~1407 is {\it not} an unusually dark system, with a group $M/L$ that is comparable to that of Fornax group ($\sim300$ $M_\odot/L_B$; Drinkwater et al.\ 2001). 

Another important observable is the baryon mass fraction. 
We determined an enclosed baryon mass fraction of 0.06 within 100 kpc for this group, which is 
much larger than previously determined by Romanowsky et al.\ (2009). 
This value is still smaller than the cosmological value of $0.17$, but comparable to other galaxy groups with similar temperatures (Dai et al.\ 2010). 
Thus, this baryon deficit is not unusual for groups. 
However, galaxy groups as a population do tend to have smaller baryon fractions than clusters (Dai et al.\ 2010). 
One explanation is that galaxy groups have shallower gravitational potentials, making them more vulnerable to AGN feedback and/or galactic winds. 
Thus, their atmospheres may have been redistributed to large radii (beyond $R_{500}$). 
There are a number of {\sl Suzaku} X-ray observatory studies of clusters reaching $R_{200}$. 
In contrast, there are only a few such studies of galaxy groups to comparable radii, 
due to their relatively lower X-ray surface brightness.      
Thus far, there are three such investigations involving poor clusters: 
Hydra A (3.0 keV; Sato et al.\ 2012), RXJ1159+5531 (2.0 keV; Humphrey et al.\ 2012) and ESO 3060170 (2.7 keV; Su et al.\ 2013). 
The enclosed baryon mass fractions in these systems are 0.23, 0.17 and 0.13, respectively. 
The surprisingly high value of Hydra A (higher than cosmic) may result from its total mass
being underestimated, due to non-thermal pressure support at large radii. 
The observations
of RXJ1159+5531 are consistent with theoretical predictions, with no baryons missing. 
Baryons may have been lost from ESO 3060170, likely due to central feedback. 
We speculate that the diversity of gas properties at $R_{200}$ should be larger among galaxy groups than galaxy clusters.

Note that if NGC~1400 were not present, the NGC~1407 group would qualify as a fossil group, 
which is defined as a group with a central dominant galaxy at least two magnitudes brighter in 
$R$-band than the second brightest galaxy within half a virial radius (Jones et al.\ 2003). 
It has been debated whether fossil groups are the end results of galaxy mergers within groups 
or are instead a transitional stage. 
Our study indicates that NGC~1400, as a newly infalling galaxy, only temporarily
disqualifies the NGC~1407 group as a fossil group.
This is consistent with some simulation work that shows that 
the gap between the brightest and the second brightest galaxy may be intermittently 
filled over time by newly infalling galaxies (von Benda-Beckmann et al.\ 2008). 
Group/cluster scaling relations  (such as X-ray luminosity -- temperature) show that fossil groups 
lie between groups and clusters in many of their properties (Khosroshahi et al.\ 2007; Miller et al.\ 2012). 
Thus, we speculate that fossil groups may be a transient phase as groups evolve into clusters 
in the hierarchical Universe. 
This also sheds light on the puzzling fact that a large fraction of fossil groups lack 
cool cores, although they are usually thought to be highly evolved, undisturbed systems (Dupke et al.\ 2010).  
We anticipate that recent mergers, as exhibited in the NGC~1407/1400 complex, may  
inhibit the monotonic growth of cool cores in many fossil groups.  

\section{\bf Summary}

The NGC~1407 group has been a subject of interest and debate due to the large radial
velocity (1200 km~s$^{-1}$) of its second brightest galaxy, NGC~1400, relative
to the central dominant galaxy NGC~1407.  
This has lead to unusually large mass-to-light ratio estimates
and suggestions that the NGC~1407 group is one of the darkest known galactic systems.
In this paper, we presented
our investigation of the NGC~1407/1400 complex with joint {\sl XMM-Newton}/{\sl Chandra} observations. We summarize our main results as follows:

1. A region of enhanced X-ray surface brightness just east of NGC~1400 is likely to be a tail of hot gas stripped from the ISM of NGC~1400. 
The metallicity of the stripped gas is $\sim$ 0.5 $Z_\odot$, larger than the metallicity of the group gas ($\lesssim 0.3~Z_\odot$). The deduced tail length is $\sim50$ kpc.

2. We studied the conditions for ram pressure stripping for NGC~1400. To have the gas in the enhanced region stripped from NGC~1400's ISM, the galaxy needs to have a relative velocity of at least 1275 km\,s$^{-1}$. This value is in excellent agreement with the possible velocity range (1200--1500 km\,s$^{-1}$) we estimated for NGC~1400 through other methods using pressure jumps and the gravitational potential.
No obvious shock feature was found in this system, which may be a consequence of NGC~1400 moving nearly in the line of sight.

3. With an {\sl XMM-Newton} pointing on the (relatively relaxed) eastern side of NGC~1407, 
we derived a hydrostatic mass  for this group of $1\times10^{13} M_\odot$ within 100 kpc. 
The total mass extrapolated to the virial radius (681 kpc) is 3.75$\times10^{13} M_\odot$. 
This provides an upper limit for the mass to light ratio of $\sim60$ $M_\odot/L_{K_\odot}$ ($\sim300$ 
$M_\odot/L_{B_\odot}$) within the virial radius. 
We thus conclude that NGC~1407 group is {\it not} an unusually dark group. 

4. We speculate that NGC~1400 recently entered the NGC~1407 group.
The presence of NGC~1400 within half a group virial radius of NGC 1407 
currently (but temporarily) disqualifies NGC~1407 from
being classified as a fossil group.
This suggests that some fossil groups may be in a transient stage, 
rather than the secular end-result of galaxy mergers within groups.

\section{Acknowledgments}
We would like to thank the anonymous referee for his/her constructive suggestions.
We thank Steve Allen, Akos Bogdan, Ka-Wah Wong, and Evan Million for useful discussions. Y.S. wishes to thank Steve Allen for his hospitality during her visit to Stanford University. We gratefully acknowledge the support from SAO/NASA {\sl Chandra/XMM} grant GO2-13163X.

\clearpage

\begin{deluxetable}{cccccccc}
\tablewidth{0pc}
\tablecaption{Observation log}
\tablehead{
\colhead{Mission}&\colhead{Name}&\colhead{Obs ID}&\colhead{Effective Exposure (ksec)}&\colhead{R.A. (deg)}&\colhead{Dec. (deg)}&\colhead{Obs Date}&\colhead{PI}}
\startdata
\multirow{2}{*}{XMM}&West &0404750101 &40, 37, 28 [m1, m2, pn] &54.91 & -18.68& 2007-2-11&B.Tully\\
&East &0679600101	&29, 29, 20 [m1, m2, pn] & 55.23 & -18.56& 2012-1-15& Y.Su\\

\hline
{Chandra}&N1400 &14033 &50 [ACIS-S]&54.92 &-18.67 & 2012-6-17& Y.Su\\
\enddata
\end{deluxetable}

\begin{deluxetable}{lllllllll}
\tablewidth{0pc}
\tablecaption{X-ray background determinations}
\tablehead{
{Method}&\colhead{S$_{CXB}$$^{a}$}&\colhead{Norm$_{CXB}$$^{b}$}&\colhead{S$_{MW}$$^{c}$}&\colhead{Norm$_{MW}$$^{d}$}&\colhead{S$_{LB}$$^{c}$}&\colhead{Norm$_{LB}$$^{d}$}&\colhead{$C/$d.o.f}}
\startdata
Best-fit  &$33.46$&$14.70^{+5.53}_{-5.08}$&$7.14$&$4.29^{+0.96}_{-0.91}$ &$4.05$&$22.07^{+3.54}_{-3.45}$& 30641.57/29064\\
{\bf Marginalized} &$33.46 $&$14.70$&$7.18$ &$4.30$&$4.08$&$22.20$&\\
\enddata
\tablecomments{ \\
a) Surface brightness for CXB emission in 0.5-2.0 keV in units of $\times 10^{-9}$ ergs s$^{-1}$ cm$^{-2}$ str$^{-1}$.\\
b) Normalization of power-law component with photon index of 1.41, in units of photons s$^{-1}$ cm$^{-2}$ keV$^{-1}$ str$^{-1}$ at 1 keV.  \\
c) Surface brightness for Galactic emissions in in 0.5-2.0 keV in the unit of $\times 10^{-9}$ ergs s$^{-1}$ cm$^{-2}$ str$^{-1}$.\\
d) Normalization is the integrated line-of-sight emission measure, 
$(1/4\pi)\int n_e n_H ds$, in units of 10$^{14}$ cm$^{-5}$ str$^{-1}$.\\
Uncertainties of surface brightnesses are proportional to these of normalizations.
} 
\end{deluxetable}

\begin{deluxetable}{ccccrr}

\tabletypesize{\small}
\tablewidth{0pc}
\tablecaption{Summary of projected fit parameters for group emission in regions 1--10}
\tablehead{
\colhead{}&\colhead{Annuli}&\colhead{Temperature}&\colhead{Abundance}&\colhead{${S_{X}}^{a}$}&
\colhead{$C$/d.o.f}\\
 & \colhead{(arcmin)}&\colhead{(keV)}&\colhead{(Z$_{\odot}$)}&\colhead{}&}
\startdata
& \phn\phn0--1.00  &$0.91^{+0.01}_{-0.01}$&$0.76^{+0.06}_{-0.04}$&$2559.48^{+408.10}_{-471.37}$& 1408.3/1400\\
&1.00--2.00  &$1.25^{+0.06}_{-0.04}$&$0.43^{+0.06}_{-0.06}$&$397.23^{+31.21}_{-29.23}$& 1714.0/1623\\
&2.00--3.15  &$1.34^{+0.02}_{-0.10}$&$0.48^{+0.20}_{-0.17}$&$151.83^{+24.50}_{-24.10}$& 1346.3/1377\\
&3.15--4.30 &$1.22^{+0.09}_{-0.09}$&$0.33^{+0.11}_{-0.08}$&$80.51^{+11.40}_{-11.43}$& 1776.2/1695\\
West &4.30--5.45  &$1.18^{+0.12}_{-0.08}$&$0.49^{+0.27}_{-0.16}$&$42.23^{+9.49}_{-11.94}$& 1875.9/1842\\
&5.45--6.60  &$1.22^{+0.07}_{-0.11}$&$0.34^{+0.15}_{-0.11}$&$40.26^{+7.79}_{-7.74}$&1951.6/1878\\
&6.60--8.00  &$1.06^{+0.09}_{-0.06}$&$0.34^{+0.11}_{-0.09}$&$44.75^{+6.96}_{-6.88}$& 2216.6/2075\\
&\phn8.00--10.50 &$1.08^{+0.07}_{-0.03}$&$0.34^{+0.10}_{-0.07}$&$41.57^{+5.88}_{-8.86}$& 2270.0/2178\\
&10.50--13.30  &$1.07^{+0.09}_{-0.08}$&$0.18^{+0.04}_{-0.03}$&$50.56^{+6.31}_{-6.29}$& 2403.6/2158\\
&13.30--16.70  &$1.04^{+0.12}_{-0.19}$&0.25 (fixed)&$13.78^{+1.46}_{-1.29}$&2320.8/2139\\
\hline
\\
& \phn\phn0--1.00  &$0.93^{+0.01}_{-0.01}$&$0.80^{+0.06}_{-0.02}$&$2408.93^{+568.94}_{-429.13}$& 1178.23/1227\\
&1.00--2.00  &$1.30^{+0.02}_{-0.01}$&$0.43^{+0.02}_{-0.04}$&$363.85^{+24.44}_{-8.14}$&1491.89/1335 \\
&2.00--3.15  &$1.46^{+0.12}_{-0.16}$&$0.37^{+0.03}_{-0.05}$&$155.07^{+5.03}_{-12.07}$& 1211.87/1285\\
&3.15--4.30 &$1.34^{+0.23}_{-0.05}$&$0.49^{+0.19}_{-0.15}$&$70.82^{+8.75}_{-28.33}$&1433.26/1393 \\
East &4.30--5.45  &$1.30^{+0.06}_{-0.07}$&$0.27^{+0.03}_{-0.11}$&$61.57^{+9.60}_{-9.60}$& 1517.23/1567\\
&5.45--6.60  &$1.32^{+0.16}_{-0.07}$&$0.30^{+0.12}_{-0.10}$&$49.50^{+9.95}_{-6.98}$&1835.24/1707\\
&6.60--8.00  &$1.26^{+0.05}_{-0.06}$&$0.32^{+0.11}_{-0.15}$&$36.54^{+7.31}_{-3.37}$& 1839.95/1833\\
&\phn8.00--10.50 &$0.95^{+0.03}_{-0.03}$&$0.26^{+0.04}_{-0.07}$&$37.18^{+2.73}_{-1.91}$&2391.51/2128 \\
&10.50--13.30  &$0.83^{+0.03}_{-0.03}$&$0.15^{+0.11}_{-0.05}$&$20.81^{+1.64}_{-3.12}$& 2228.67/2110\\
&13.30--16.70  &$0.86^{+0.05}_{-0.02}$&0.25 (fixed)&$5.55^{+1.81}_{-1.25}$&2118.57/2098\\
\enddata
\tablecomments{ 
a) Surface brightness of group emission (0.5-2.0 keV) in units of 
$10^{-9}$ erg s$^{-1}$ cm$^{-2}$ sr$^{-1}$.
}
\end{deluxetable}

\begin{deluxetable}{ccccc}
\tablewidth{0pc}
\tablecaption{Summary of deprojected fit parameters for group emission in regions 1--10}
\tablehead{
\colhead{Annuli}&\colhead{Temperature}&\colhead{Abundance}&\colhead{$n_{\rm e}$}&\colhead{$C$/d.o.f}\\
 \colhead{(arcmin)}&\colhead{(keV)}&\colhead{(Z$_{\odot}$)}&\colhead{($\times$\,10$^{-3}$\,cm$^{-3}$)}&}
\startdata
\phn\phn0--1.00  &$0.88^{+0.01}_{-0.01}$ &$0.86^{+0.02}_{-0.03}$&$8.27^{+0.09}_{-0.09}$& \multirow{10}{*}{17831.8/17495}\\
1.00--2.00  &$1.27^{+0.02}_{-0.09}$&\multirow{5}{*}{$\scalebox{2.0}{\Bigg\}}$$0.40^{+0.02}_{-0.02}$}&$3.70^{+0.08}_{-0.07}$& \\
2.00--3.15  &\multirow{2}{*}{$\Big\}$$1.54^{+0.10}_{-0.10}$}&&$1.30^{+0.05}_{-0.05}$& \\
3.15--4.30 &&&$0.55^{+0.06}_{-0.07}$& \\
4.30--5.45  &\multirow{2}{*}{$\Big\}$$1.28^{+0.06}_{-0.06}$}&&$0.51^{+0.04}_{-0.04}$& \\
5.45--6.60  &&&$0.39^{+0.03}_{-0.03}$&\\
6.60--8.00  &\multirow{2}{*}{$\Big\}$$1.08^{+0.04}_{-0.02}$}&\multirow{4}{*}{$\scalebox{1.6}{\Bigg\}}$$0.34^{+0.02}_{-0.02}$}&\multirow{2}{*}{$\Big\}$$0.33^{+0.01}_{-0.01}$}& \\
\phn8.00--10.50 &&& \\
10.50--13.30  &$1.08^{+0.08}_{-0.05}$&&$0.19^{+0.01}_{-0.01}$& \\
13.30--16.70  &$0.97^{+0.12}_{-0.12}$&&$0.09^{+0.01}_{-0.01}$&\\

\enddata
\tablecomments{ 
To ensure the stability of the deprojected spectral fit, we had to link the temperatures of regions 3 and 4,
regions 5 and 6, and regions 7 and 8. We also needed to tie the electron densities of regions 7 and  8 
together. The metal abundances of regions 2--6 and regions 7--10 were linked together respectively.}

\end{deluxetable}

\begin{deluxetable}{ccccccccc}
\tabletypesize{\small}
\tablewidth{0pc}
\tablecaption{Summary of parameters for NGC~1400 and the enhanced region}
\tablehead{
\colhead{Name}&\colhead{Aperture}&\colhead{$T$}&\colhead{${L_{\rm X}}^a$}&\colhead{$L_{\rm K}$}&\colhead{Fe}&\colhead{$C$/d.o.f}\\
 & \colhead{(arcmin)}&\colhead{(keV)}&\colhead{$(10^{39}$ erg~s$^{-1})$}&\colhead{$(10^{10} \rm L_{K_\odot})$}&\colhead{}&}
\startdata
NGC~1400& 0.98  &$0.68^{+0.06}_{-0.05}$&$3.1$&1.1&0.7 (fixed)&1157.78/1242\\
enhanced region&1.40&$0.81^{+0.03}_{-0.03}$&$4.3$&0& 0.49$^{+0.33}_{-0.14}$&1419.58/1600\\
\enddata
\tablecomments{a) X-ray luminosity of hot gas in 0.1--2.0 keV.}
\end{deluxetable}

\clearpage

\clearpage

\clearpage

\clearpage

\end{document}